\documentclass[10pt, aps,pra,floatfix,twocolumn,amsfonts,amsmath,amssymb,tightenlines,nofootinbib, longbibliography]{revtex4-1}

\usepackage[utf8]{inputenc}
\usepackage[T1]{fontenc}
\usepackage[UKenglish]{babel}
\usepackage{graphicx}
\usepackage{amsmath,amssymb,amsthm}
\usepackage{color}
\usepackage{psfrag}
\usepackage{ifsym}
\usepackage{epstopdf}
\usepackage{dsfont}
\usepackage{multirow} 
\usepackage{paralist}
\usepackage{enumitem}
\usepackage[normalem]{ulem}
\usepackage{units}
\usepackage{tabularx}
\usepackage{bbm}
\usepackage[utf8]{inputenc}
\usepackage[T1]{fontenc}
\usepackage{lmodern}
\usepackage[colorlinks]{hyperref}
\usepackage{xspace}
\usepackage{mathtools}
\usepackage{dsfont}
\usepackage{lipsum}
\usepackage{siunitx}
\usepackage{pgf,tikz}
\usepackage{pgfplots}
\usepackage{footnote}
\usepackage{soul,xcolor}
\setstcolor{blue}

\DeclareMathOperator{\Tr}{Tr}

\newcommand{\ket}[1] {| #1 \rangle}

\newcommand{\ketbra}[2]{ | #1 \rangle\!\langle #2 |}

\newcommand{\id}{\mathds{1}}

\newcommand{\chan}[2]{{T}^{#1  #2 _I}}

\newcommand{\ba}{\begin{eqnarray}}
\newcommand{\ea}{\end{eqnarray}}
\newcommand{\be}{\begin{equation}}
\newcommand{\ee}{\end{equation}}
\newcommand{\bs}{\begin{split}}
\newcommand{\es}{\end{split}}
\newcommand{\bef}{\begin{figure}[!h]}
\newcommand{\eef}{\end{figure}}

\newtheorem*{lem*}{Lemma}

\hypersetup{pdftitle={{A quantum causal discovery algorithm}}}

\begin{document}

\title{A quantum causal discovery algorithm}

\author{Christina Giarmatzi$^{1,2}$ and Fabio Costa$^{1}$}
\affiliation{$^{1}$Centre for Engineered Quantum Systems, $^{2}$Centre for Quantum Computer and Communication Technology, School of Mathematics and Physics, University of Queensland, Brisbane, QLD 4072, Australia}

\begin{abstract}
Finding a causal model for a set of classical variables is now a well-established task---but what about the quantum equivalent? Even the notion of a quantum causal model is controversial. Here, we present a causal discovery algorithm for quantum systems.  The input to the algorithm is a process matrix describing correlations between quantum events. Its output consists of different levels of information about the underlying causal model. Our algorithm determines whether the process is causally ordered by grouping the events into causally-ordered non-signaling sets.  It detects if all relevant common causes are included in the process, which we label Markovian, or alternatively if some causal relations are mediated through some external memory. For a Markovian process,  it outputs a causal model, namely the causal relations and the corresponding mechanisms, represented as quantum states and channels. Our algorithm provides a first step towards more general methods for quantum causal discovery.
\end{abstract}
\maketitle

\section{Introduction}
The discovery of causal relations is a basic and universal task across all scientific disciplines. The very nature of causal relations, however, has been a long-standing subject of controversies with the central question being what, if anything, distinguishes causation from correlation.

It is only recently that a rigorous framework for causal discovery has been developed \cite{Pearlbook, spirtes2000causation}. Its core ingredients are \emph{causal mechanisms} that are responsible for correlations between observed \emph{events}, with the possibility of external \emph{interventions} on the events. The possibility of interventions is what provides an empirically well-defined notion of causation, distinct from correlation: an event $A$ is a cause for an event $B$ if an intervention on $A$ results in a change in the observed statistics of $B$. A causal model is typically defined as a set of direct-cause relations and a quantitative description of the corresponding causal mechanisms. The causal relations are represented as arrows in a graph and the causal mechanisms are usually described in terms of transition probabilities (Figure~\ref{fig:cause_effect2}).

\bef
\includegraphics[width=.8\linewidth]{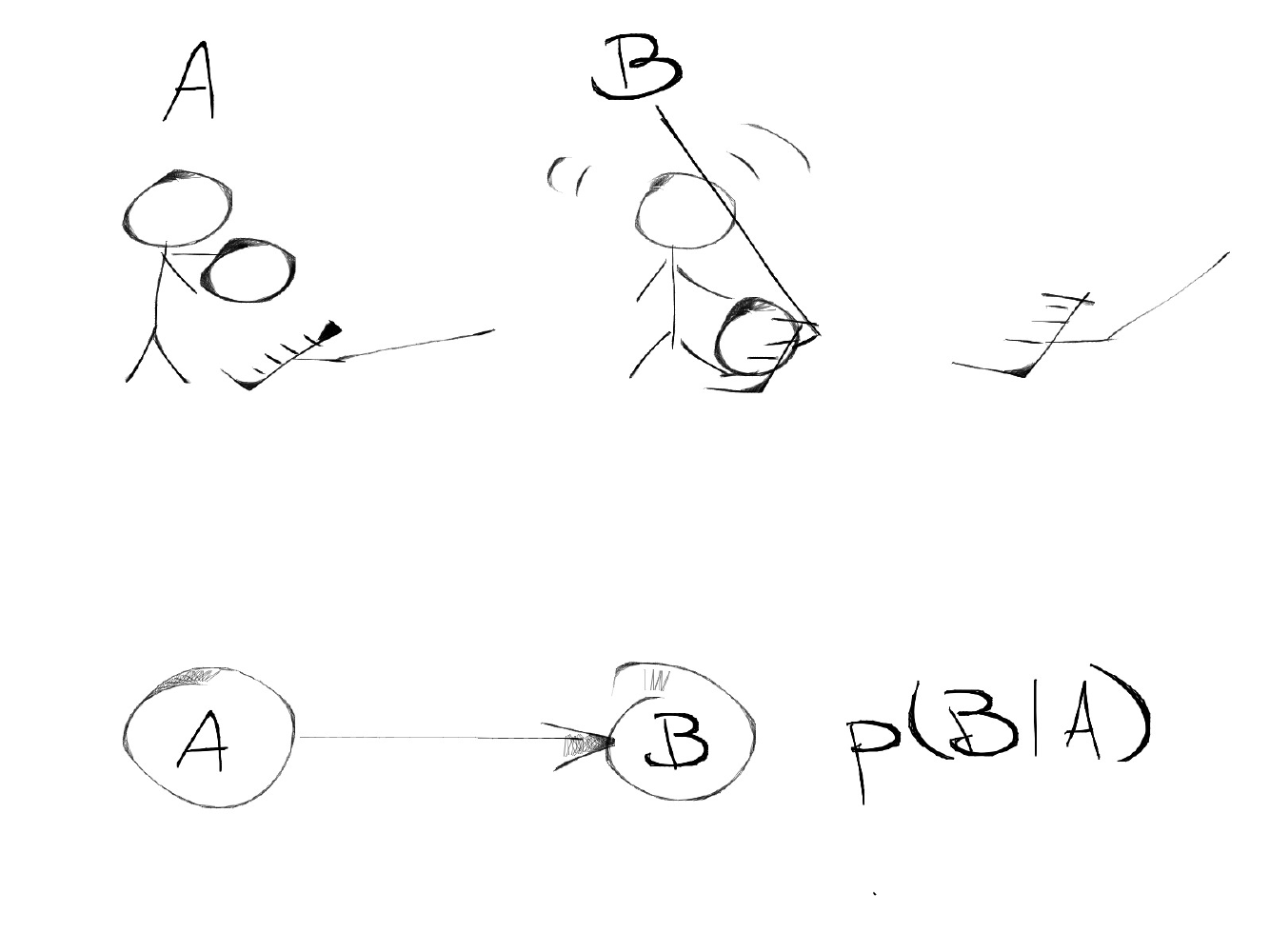}
\caption{{\bf Causal relations: }An example of a causal relation and its representation in a graph.}
\label{fig:cause_effect2}
\eef

Among the most important achievements of causal models is the development of algorithms for causal discovery. The objective of such algorithms is to infer a causal model based on observational and interventional data. Such algorithms have found countless applications and constitute one of the backbones in the rising field of machine learning.

It is a natural question whether similar algorithms can be developed for quantum systems. In simple quantum experiments, causal relations are typically known and well under control. However, the fast growth of quantum technology goes towards the development of networks of increasing size and complexity. Hence, appropriate tools to recover causal relations might become necessary for the functioning of large, distributed quantum networks, as it is already the case for classical ones \cite{Lamport1978}. Causal discovery might further detect the presence of ``hidden common causes'', namely external sources of correlations that might introduce systematic errors. Finally, from a foundational perspective, the possibility of discovering causal relations from empirical data opens the possibility to recover causal structure from more fundamental primitives.

Classical causal discovery algorithms, however, fail to discover causal relations in quantum experiments~\cite{woodlesson2012}. A considerable effort has been recently devoted to solve this tension and transfer causal modeling tools to the quantum domain \cite{tucciquantum1995, Leifer2006, Laskey2007, Leifer2013, cavalcanti2014modifications,
fritzbeyond2015, Henson2015, pienaar2014graph, chavesinformation2015, ried2015quantum}, leading to the formulation of a quantum causal modeling framework \cite{costa2016, Allen2016}. (See Refs.~\cite{Shrapnel2015, Shrapnel2016} for a broader philosophical context.)

Here we introduce a first algorithm for the discovery of causal relations in quantum systems. 
The starting point of the algorithm is a description of a quantum experiment (or ``process'') that makes no prior assumption on the causal relations or temporal order between events~\cite{oreshkov12}. Given such a description, encoded in a \emph{process matrix}, the algorithm extracts different levels of causal information for the events in the experiments. It determines whether or not they are \emph{causally ordered}, namely whether they can be organized in a sequence where later events cannot influence earlier ones. If a causal order exists, the algorithm finds if all common causes are modeled as events in the process matrix---a property expressed by the condition of \emph{Markovianity}, as defined in Ref.~\cite{costa2016}. If the process is Markovian, the algorithm outputs a causal model for it: a causal structure (depicted as arrows connecting events) together with a list of quantum channels and states that generate the process.

The complexity of our algorithm scales quadratically with the number of events, although the size of the problem itself (the dimension of the process matrix) is exponential. This suggests that the algorithm can be used efficiently given an efficient encoding of the input to the code. We further comment on possible extensions of the algorithm to deal with processes that are not markovian, not causally ordered, or that follow a different definition of markovianity~\cite{Allen2016}. We provide the implementation of the algorithm, written on MatLab, together with the required functions, some examples, and a Manual~\cite{discovery_code2017}. The code uses some functions developed by Toby Cubitt ~\cite{Cubitt}.

\section{Quantum causal models}
\subsection{Process framework}
We will use a formulation of quantum mechanics that can assign probabilities to quantum events with no prior knowledge of their causal relations~\cite{oreshkov12}. This formulation is based on the ``combs'' formalism for quantum networks~\cite{chiribella09b}, with the main difference that the causal order between events is not assigned in advance. 

In this framework, a {\it quantum event} $A$ can be thought to be performed by a party inside a closed laboratory (Figure~\ref{fig:lab_event})---which is associated with an input and an output Hilbert space, ${\cal H}^{A_I}$ and ${\cal H}^{A_O}$ respectively---and is represented by a completely positive (CP) map ${\cal M}^{A_I\rightarrow A_O} : {\cal L}({\cal H}^{A_I}) \rightarrow {\cal L}({\cal H}^{A_O})$, where ${\cal L}({\cal H}^{S})$ is the space of linear operators over the Hilbert space of system $S$. A {\it quantum instrument} is the collection of CP maps ${\cal J}^A = \{ {\cal M}^A\}$, such that $\underset{{\cal M}^A\in {\cal J}^A}{\sum} {\cal M}^A$ is a CP and trace-preserving (CPTP) map. 

\bef
\includegraphics[width=.9\linewidth]{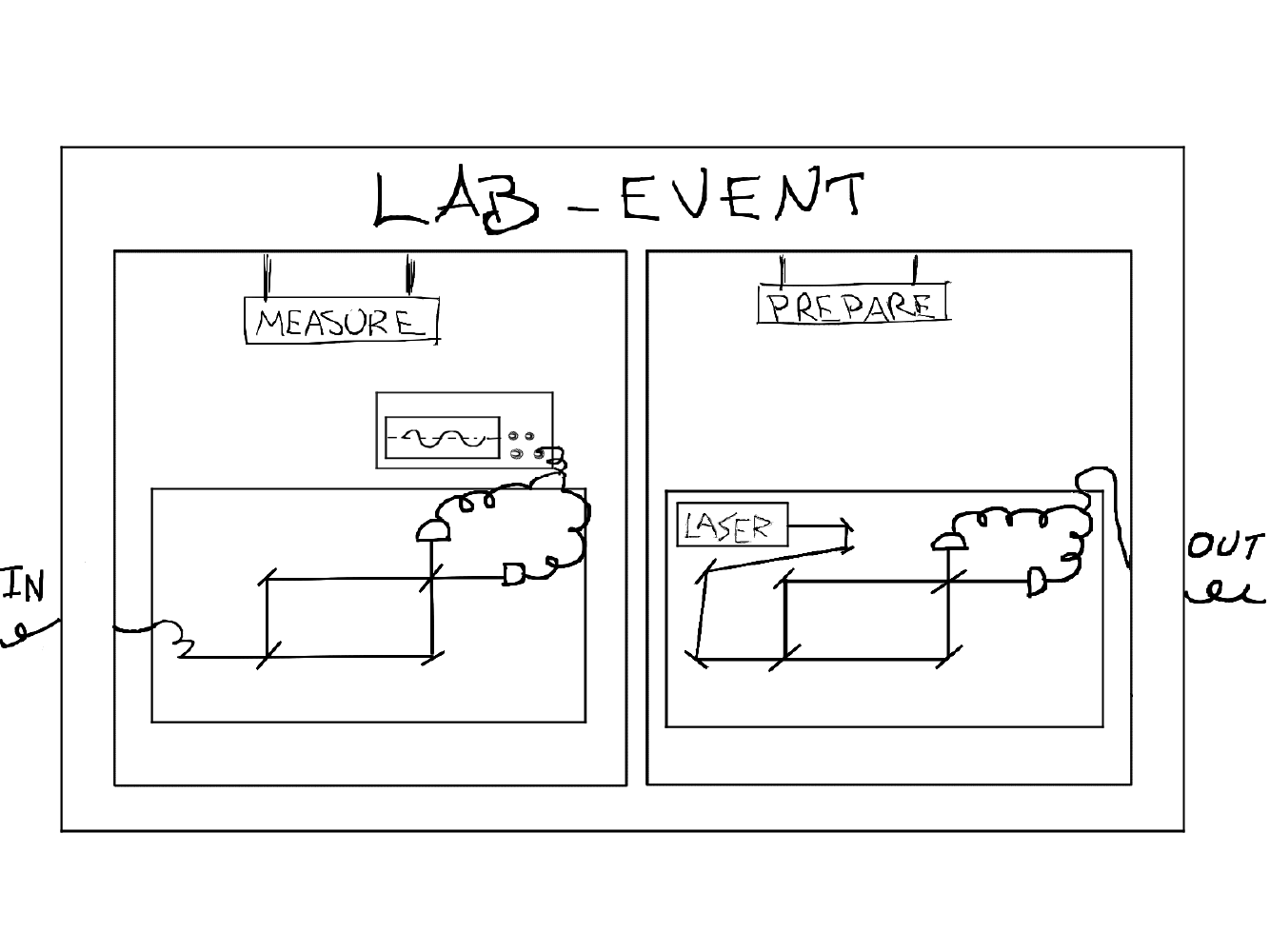}
\caption{{\bf Lab-event:} A picture of a quantum event, consisting of a measurement stage of the input system and a preparation stage for the output system. It may also be simply a unitary transformation.}
\label{fig:lab_event}
\eef

It was found that, for a set of parties $\{A^1, \cdots, A^n\}$, the joint probability of their CP maps to be realized, given their instruments, is a function of their maps and some matrix that mediates their correlations:

\ba
p({\cal M}^{A^1}, \cdots, {\cal M}^{A^n} | {\cal J}^{A^1},\cdots, {\cal J}^{A^n}) = \nonumber \\ 
\Tr[W^{A^1_IA^1_O \cdots A^n_IA^n_O} (M^{A^1_IA^1_O}\otimes \cdots \otimes M^{A^n_IA^n_O})].
\label{eq:prob}
\ea
Using a version of the Choi-Jamiolkovsky (CJ) isomorphism \cite{jamio72, Choi1975}, the CJ matrix $M^{A_IA_O}\in {\cal L}({\cal H}^{A_I}\otimes{\cal H}^{A_I})$, isomorphic to a CP map ${\cal M}^{A} : {\cal L}({\cal H}^{A_I}) \rightarrow {\cal L}({\cal H}^{A_O})$ is defined as $M^{A_IA_O} := [{\cal I} \otimes {\cal M}(\ketbra{\phi^+}{\phi^+})]^T$, where $\cal I$ is the identity map, $\ket{\phi^+} = \sum_{j=1}^{d_{A_I}}\ket{jj} \in {\cal H}^{A_I} \otimes {\cal H}^{A_I}$, $\{\ket{j}\}^{d_{A_I}}_{j=1}$ is an orthonormal basis on ${\cal H}^{A_I}$ and ${ T}$ denotes matrix transposition in that basis and some basis of ${\cal H}^{A_O}$. Finally, $W^{A^1_IA^1_O, \cdots, A^n_IA^n_O} \in {\cal L}({\cal H}^{A^1_I}\otimes{\cal H}^{A^1_O}\otimes \cdots \otimes{\cal H}^{A^n_I}\otimes{\cal H}^{A^n_O})$ is a {positive semi-definite} matrix that lives on the combined Hilbert space of all input and output systems of the parties and is called {\it process matrix}. Equation~\eqref{eq:prob} can be seen as a generalization of the Born rule, and the process matrix as a generalization of the quantum state, as it is the resource that allows calculating joint probabilities for all possible events. Just as the Born rule is the only non-contextual probability assignment for POVM measurements~\cite{caves2004}, Equation~\eqref{eq:prob} is the only non-contextual probability rule for CP maps~\cite{shrapnel2017updating}.

Here we are interested in situations where causal relations define a partial order, which we call \emph{causal order}. We identify causal relations with the possibility of signaling: if the probability of obtaining an outcome in laboratory $B$ can depend on the settings in laboratory $A$, we say that $A$ \emph{causally precedes} $B$, and write $A\prec B$. (We write $A||B$ if no signaling is possible and say $A$ and $B$ are \emph{causally independent}.)  The process matrices that define a causal order between the events are called {\it causally ordered}.

\subsection{From mathematical to graphical representation}
The causal structure encoded in the process matrix can be represented by a {\it Directed Acyclic Graph (DAG)}. A directed graph is a pair ${\cal G = \langle V,E\rangle}$, where ${\cal V}= \{V_1, ...,V_n\}$ is a set of {\it vertices} (or nodes) and ${\cal E \subset V \times V}$ is a set of ordered pairs of vertices, representing {\it directed edges}. A {\it directed path} is a sequence of directed edges where, for each edge, the second vertex is the first one in the next edge. Figure~\ref{fig:subfigures} ($\alpha$) shows a directed path from $V_1$ to $V_3$. A {\it directed cycle} is a directed path that ends up in a vertex already used by the path, as shown in Figure~\ref{fig:subfigures} ($\beta$). A DAG is a directed graph with no directed cycles, as shown in Figure~\ref{fig:subfigures} ($\gamma$). We refer to edges as \emph{causal arrows}.

\bef
\includegraphics[width=\linewidth]{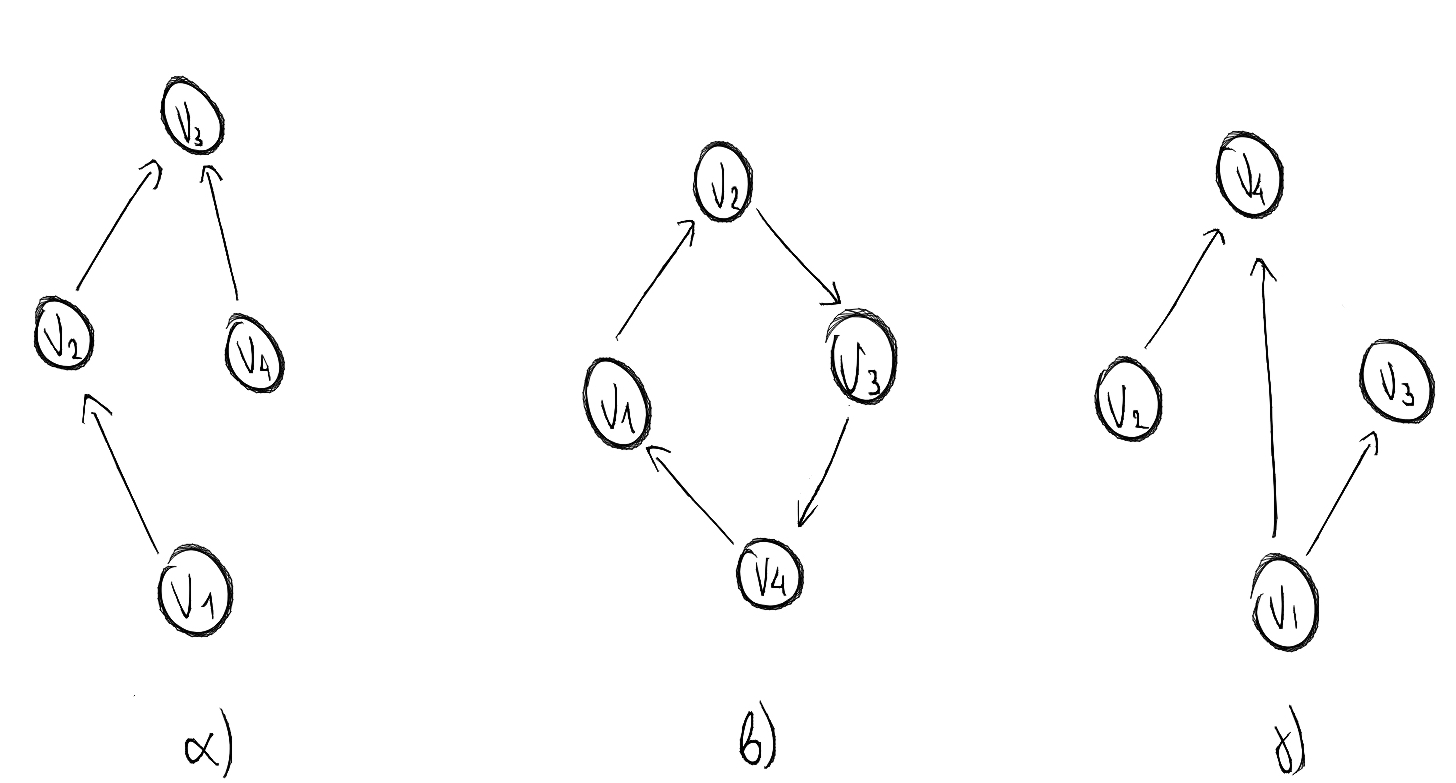}
    \caption{{\bf Examples:} Figure ($\alpha$) shows a DAG with a directed path from $V_1$ to $V_3$, ($\beta$) an example of a directed cycle and ($\gamma$) another example of a DAG.}   \label{fig:subfigures}
\eef

Following Ref.~\cite{costa2016}, we define a quantum causal model by associating a specific type of process matrix to a DAG. We associate a party, with input and output spaces, to each node of the DAG. If the node has more than one outgoing arrow, the output space is composed of subsystems, with one subsystem for each arrow. We refer to them as \emph{output subsystems}. We define the \emph{parent space} $\Gamma^A$ of a node $A$ as the tensor product of all output subsystems associated with an arrow ending in $A$. A \emph{Markov quantum causal model} is then defined by a collection of quantum channels, one for each node $A$, connecting the parent space of $A$ to its input space.

Now let us see how a process matrix whose causal structure is represented by a DAG looks like. It will be a tensor product of three types of factors: input states for the set of parties with no incoming arrow in the DAG, channels connecting each input system of a remaining party with its parent space, and finally the identity matrix $\id$ for the output systems of the set of parties with no outgoing arrows in the DAG. For example, if $\{F^1, F^2, ..., M^1, M^2, ... L^1, L^2, ... \}$ is a set of parties where $F$, $M$ and $L$ is the label for the three set of parties described above (first, middle, and last), respectively, then their process matrix would be
\be
W^{F^1_IF^1_O...} = \rho_1^{F^1_I}\otimes \rho_2^{F^2_I}\otimes \cdots \chan{\Gamma^{M^1}}{M^1}\otimes \chan{\Gamma^{M^2}}{M^2}\otimes ...  \id^{L^1_OL^2_O...},
\label{eq:markovW}
\ee 
where $\chan{\Gamma^{M^j}}{M^j}$ is a matrix representing a CPTP map ${\cal T}$ from $\Gamma^{M^j}$ to $M^j_I$ via the isomorphism\footnote{This isomorphism is the same as the one used to describe the CP maps of the parties, but without transposition.} $\chan{\Gamma^{M^j}}{M^j} := {\cal I} \otimes {\cal T}(\ketbra{\phi^+}{\phi^+}) \in {\cal H}^{\Gamma^{M^j}}\otimes {\cal H}^{M^j_I}$. 
From now on we identify a channel with its matrix representation.
A representation of Markovian processes as Equation~\eqref{eq:markovW} is also employed in the study of open quantum systems~\cite{pollockcomplete2015}.

The above condition for the causal structure of the process matrix to be described by a DAG is a quantum generalisation of the Markov condition for classical variables and so it can be called the quantum Markov condition~\cite{costa2016}. (We will comment below on a slightly different possible definition~\cite{Allen2016}.) Such a process matrix is also causally ordered, with a partial order defined by the DAG. However, the class of causally ordered process matrices is strictly broader than Markovian ones, and they are represented by quantum combs~\cite{chiribella09b}. {As we will see later, causally ordered processes that are not Markovian can be understood as processes involving correlations with some unobserved systems---called `latent' variables. The algorithm we present here detects whether a process matrix is causally ordered and if it is, it outputs the causal order of sets of parties that are causally independent. It further detects Markovianity and for a Markovian process it outputs the DAG associated with the process matrix. We discuss in section ``Non-Markovian processes'' possible extensions of the algorithm that could output a DAG for a non-Markovian process.}

\section{quantum causal discovery}
The input to the code is a process matrix, which can be obtained from experimental data. The procedure is similar to quantum state tomography: one can reconstruct the process matrix given the probabilities arising from informationally instruments~\cite{costa2016}.

\subsection{The linear constraints}
A process matrix of the form of Equation~\eqref{eq:markovW} satisfies a set of linear constraints. This set identifies a DAG---in fact, each constraint corresponds to a particular element in the DAG. There are two types of constraints.

{\bf Open output:} A party $A$ has an \emph{open output} when in the process matrix $W$ there is an identity matrix on the corresponding output system ${A_O}$. This translates to the following linear constraint: 
\be
\tilde{\id} ^{A_O} \otimes \Tr_{A_O} W= W
\label{eq:con1}
\ee
where $\tilde{\id} ^{A_O} = \id^{A_O}/d_{A_O}$ and $d_{A_O}$ is the dimension of the system $A_O$.
When this condition is satisfied, the party $A$ cannot signal to any party and is considered \emph{last}. In the case where the output system of the party is decomposed into {output} subsystems $A_{O_i}$, $i = 1, \cdots, n$, then the corresponding identity matrix in the process matrix lives on the Hilbert space of that output subsystem $A_{O_i}$. We also call this subsystem \emph{open} and the linear constraint is 
\be
\tilde{\id} ^{A_{O_i}} \otimes \Tr_{A_{O_i}} W= W.
\label{eq:con1i}
\ee

{\bf Channel:} 
A quantum channel between the input of a party $A$ and its parents space $\Gamma^A$ is represented by a factor $\chan{\Gamma^A}{A}$ in the process matrix, as we have already mentioned. It is a positive matrix that lives on the tensor product of the Hilbert spaces of the output and input systems involved, and has the property that upon tracing out the output of the channel (the input of $A$) what remains is identity on the input (the space of output systems $\Gamma^A$):
\be
\Tr_{A_I} \chan{\Gamma^A}{A} = \id^{\Gamma^A}.
\label{eq:channel}
\ee
This property is necessary and sufficient for the channel to be trace preserving and we use it to discover channels in the process matrix: we trace out the input of $A$, $A_I$, and we check whether in the remaining process matrix there is identity on the output system of a given party, say, $B$. This describes a linear constraint that a process matrix satisfies when there is a channel from the output of $B$ to the input of $A$.

\be
\tilde{\id}^{B_O} \otimes \Tr_{B_O} (\Tr_{A_I} W) = \Tr_{A_I} W
\label{eq:con2}
\ee
If the output of party $B$ is decomposed into subsystems, then we use the above constraint for each subsystem separately, by replacing $B_O$ with every output subsystem $B_{O_i}$.
\be
\tilde{\id}^{B_{O_i}} \otimes \Tr_{B_{O_i}} (\Tr_{A_I} W) = \Tr_{A_I} W
\label{eq:con2i}
\ee
{Note that {conditions \eqref{eq:con2} and \eqref{eq:con2i} are also satisfied for open systems and subsystems, respectively.}
However, the algorithm checks {conditions \eqref{eq:con1} and \eqref{eq:con1i} first and does not consider again those (sub)system that have been tagged ``open''. Therefore, it will not associate a channel to open (sub)systems.}}The maximal set of output systems and subsystems for which conditions~\eqref{eq:con2} and~\eqref{eq:con2i} holds is the parent space of $A$, $\Gamma^A$.

In the concrete implementation of the algorithm, the above equalities are tested up to some precision defined by a small number $\epsilon$, which can be adjusted depending on the working precision. {When testing examples generated numerically, this permits one to take into account the}
different numerical rounding of {non-integer} numbers that {might otherwise} lead to errors---{for example, }$\sqrt{2}$ {defined up to some digit will be different to $\sqrt{2}^2/\sqrt{2}$ as the rounding of the last digit in different steps of the calculation will cause a different result. Naturally, the number $\epsilon$ can {also} be adjusted to account for experimental inaccuracies{, when the process matrix is obtained from experimental data}.}

\subsection{The code}
The causal discovery code subjects the process matrix to the above types of linear constraints and the set of them that are satisfied define the DAG. 

The code takes as input: the number of parties, the dimension of each input system, output system, output subsystem, and the process matrix. {The code assumes that the process matrix is positive semi-definite. Hence, its output is meaningful only if this assumption is satisfied.}

Briefly the procedure of causal discovery goes as follows: First the code identifies and traces out any open output subsystems. Then it determines whether the process matrix is causally ordered. If it is, it outputs a possible causal order and proceeds to determine if the process is Markovian. For a Markovian process, it outputs the DAG, and the represented mechanisms. Below we expand on these three stages. {In the Appendix, we show how the code works using an example of a $4$-partite process matrix. We present the causal information extracted in the different stages for that example, as well as the final output of the code.}

\paragraph{Tracing out open output subsystems:}
The code checks each output subsystem to identify if it is open, using the linear constraint in Equation~\eqref{eq:con1i}. Each found open subsystem is traced out from the process matrix, keeping track of the label of the party and the label of the subsystem, for example, subsystem 3 of party 2. Keeping track of open subsystems is what allows the algorithm to find a \emph{minimal} DAG, namely without extra arrows, as discussed below.

\paragraph{Checking if W is causally ordered:}
Let us call a {\it non-signaling set}, a set of parties that are causally independent, namely that cannot signal to each other. A non-signaling set is \emph{maximal} if it is not a proper subset of another non-signaling set. The first output of the algorithm is all the maximal non-signaling sets and their causal order. This is done through the linear constraint that detects open output systems, in Equation~\eqref{eq:con1}. The set of parties whose output systems satisfy the constraint is labeled as \emph{last set}. Note that the constraint has to be satisfied by the whole output system and not only by some subsystems. To determine the next set, the \emph{second last}, the code traces out the last set from the process matrix and, using the same constraint, it identifies the new last set, and so on. Note that the partition into maximal non-signaling sets does not uniquely identify the partial order of the parties, in the sense that it is not guaranteed that parties in different non-signaling sets can signal to each other. What is guaranteed is that at least one party from a set $\cal X$ can signal to at least one party in a succeeding set $\cal Y$ (Figure~\ref{fig:sets}). Note also that the partition into maximal non-signaling sets is not unique, much like a foliation of space-time into space-like hypersurfaces.
\bef
\includegraphics[width=.8\linewidth]{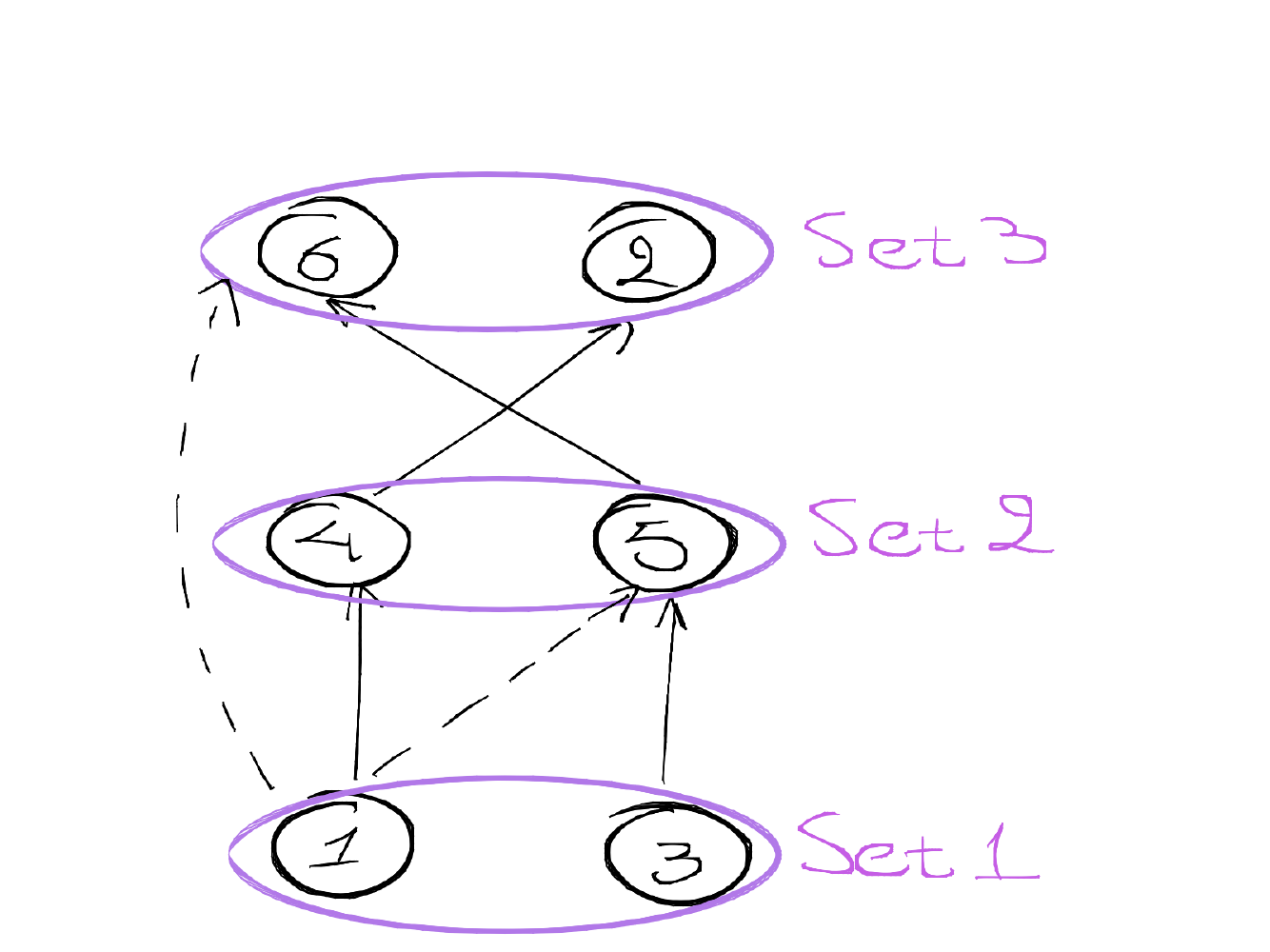}
\caption[]{{\bf Maximal non-signaling sets: }The first output of the code is a grouping of the parties into maximal non-signaling sets. The solid arrows represent a DAG compatible with this grouping. Not all parties in different sets are linked by causal arrows; the dashed arrows are examples of these missing links.}
\label{fig:sets}
\eef

The process matrix is causally ordered if and only if the algorithm succeeds in grouping all parties in maximal non-signaling sets. This is because, given the non-signaling sets, we can define a total order among the parties by adding arbitrary order relations among members of each set. For example, we can order the parties in different time steps where: when $A\prec B$, $A$ occurs at a time before $B$ and, when $A||B$, we pick an arbitrary time ordering (Figure~\ref{fig:causal_order}). With the parties ordered in this way, the process matrix satisfies the condition defining a quantum comb \cite{chiribella09b}. This is a recursive version of Equation~\eqref{eq:con1}, that holds for the output of each system after all systems that come after it are traced out. A central result in the theory of quantum networks is that, whenever this condition holds, the corresponding process can be realized as a channel with memory~\cite{gutoski06, Kretschmann2005, chiribella09b}. Thus, this part of the algorithm determines whether the input process matrix has a physical realisation as a causally ordered process.

\bef
\includegraphics[width=.9\linewidth]{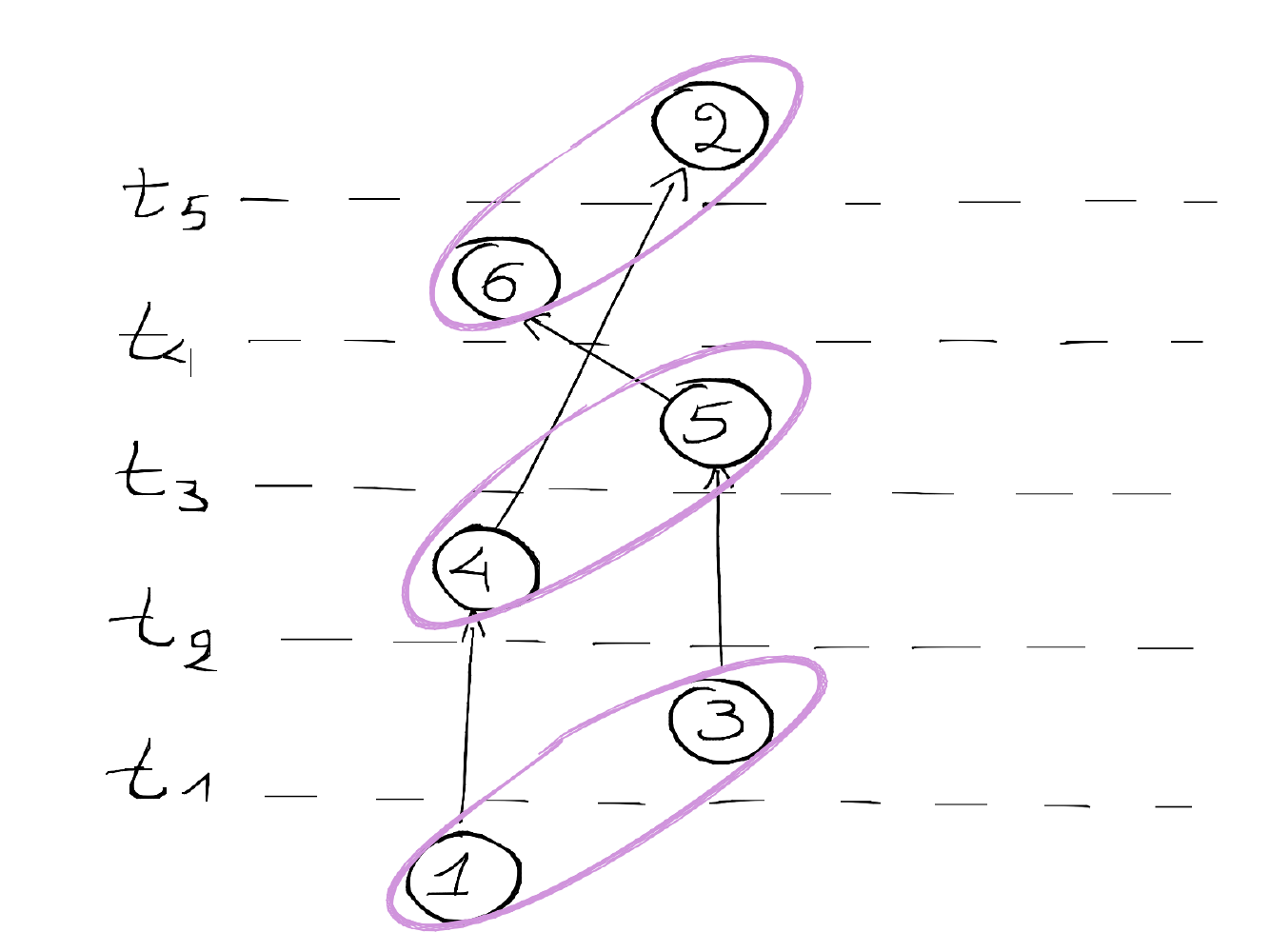}
\caption[]{{\bf Total causal order: }Starting from the DAG in Figure~\ref{fig:sets}, we can order all events in time, obtaining a total order of the parties, by putting an arbitrary order between parties in the same non-signaling set.}
\label{fig:causal_order}
\eef

\paragraph{Causal discovery and Markovianity:}
After the algorithm has traced out all open output subsystems and has established the maximal non-signaling sets of the parties, it is time to determine the DAG. The algorithm checks all possible causal arrows{---compatible with the previously found causal order---}between pairs of an input system of a party and an output system of another party, using Equation~\eqref{eq:con2}. If a party's output system is divided into subsystems, then each subsystem is checked using the linear constraint in Equation~\eqref{eq:con2i}. {To check if these constraints are satisfied, the algorithm has to check each possible one individually. In particular, for each input system that is traced out, it checks whether the constraint holds for each output system or subsystem that has not been associated yet with a causal arrow.} Every time the constraint is satisfied, an causal arrow is associated with the corresponding systems and the output system or subsystem is marked as used and is not being checked again. The collection of all output systems and subsystems that satisfy the constraints for a single input system of a party $A$ uniquely identifies the parent space of $A$, $\Gamma^A$. 
Figure~\ref{fig:stages} shows the information input to the code, and the output information that obtains during the three staged described above.

\bef
\includegraphics[width=\linewidth]{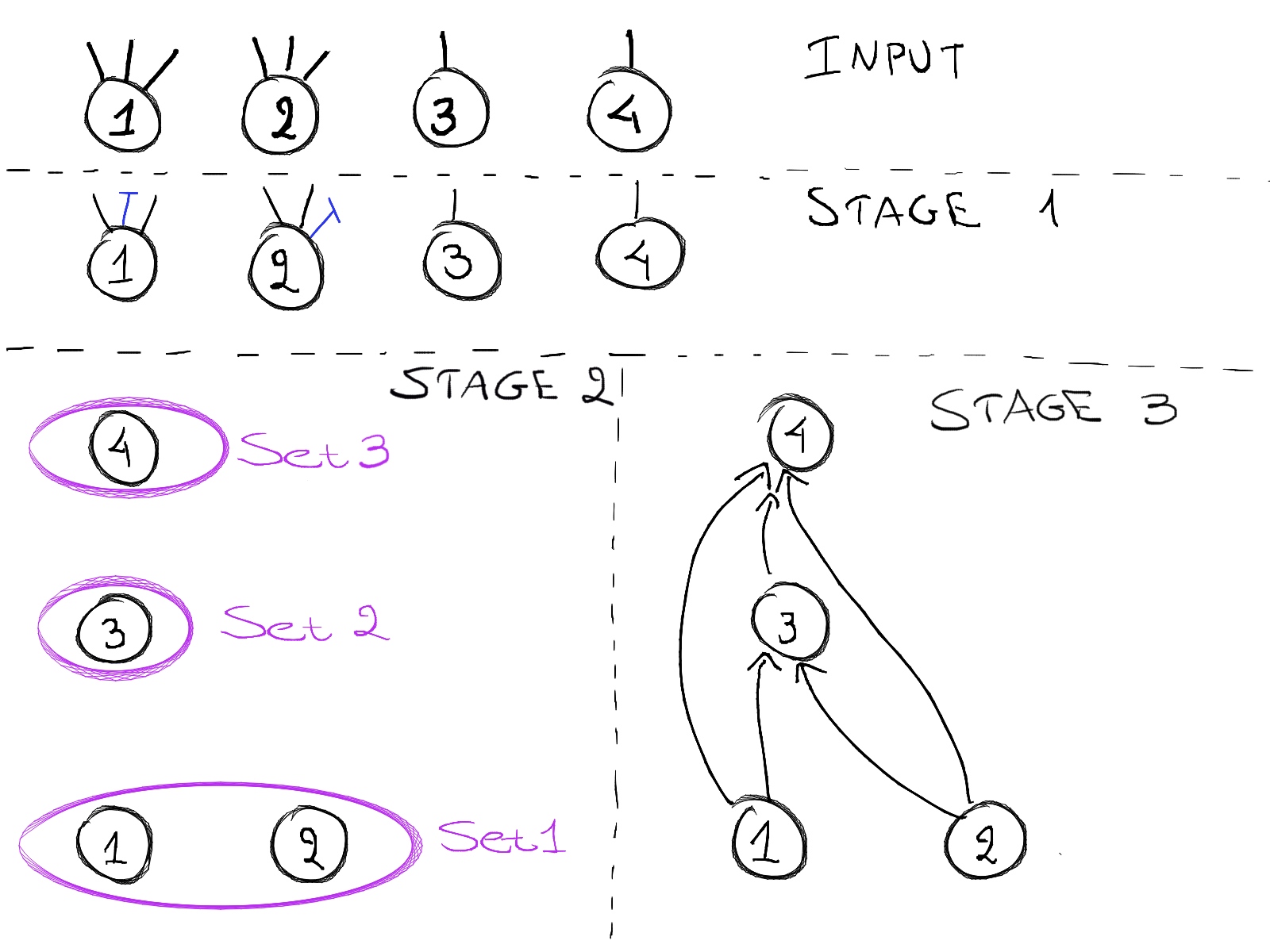}
\caption[]{{\bf Stages of the algorithm: }{As part of the input, we depict the parties and the information about their output systems and subsystems. Stage one traces out the open output subsystems, depicted in blue. The rest of the systems in black are output systems and subsystems. For a causally ordered process, stage two groups the parties into maximal non-signaling sets. For a Markovian process, stage three provides the causal model. There is no arrow for the output system of party 4 as it is last.}}
\label{fig:stages}
\eef

At this stage, the code outputs the DAG if the process is Markovian, namely if the process matrix is of the form of Equation~\eqref{eq:markovW}. To determine this, the code constructs a test-matrix that is Markovian with respect to the found DAG: it contains all (and only) the factors as in Equation~\eqref{eq:markovW} that correspond to the elements of the DAG. There are three kinds of these elements: first parties, causal arrows, and last parties; the corresponding terms on the process matrix are input states for the first parties, channels that live on the input and output systems and subsystems of the associated parties, and identity matrices on the output system of the last parties, respectively. To construct the test process matrix, these factors are extracted from the original process matrix by tracing out all systems except from the desired ones. If the process is Markovian, then the test-matrix will be equal to the original process matrix that was input to the code. 

\subsection{Minimality}
The code is guaranteed to give a unique and minimal DAG for a Markovian process. A process matrix is said to be Markov with respect to the DAG if every channel (found by Equation~\eqref{eq:con2} and Equation~\eqref{eq:con2i}) in the process matrix is represented by an arrow in the DAG. However, a $W$ can be Markov to more than one DAG---some DAGs will have arrows allowed by the causal order but there is no actual channel in the $W$ corresponding to this arrow. In other words, a $W$ can be in the tensor product form \eqref{eq:markovW}, but with some factor of the form $\chan{\Gamma^{M}}{M}=\id^{\Gamma^M}\otimes \rho^{M_I}$, for some normalized density matrix $\rho$. This represents a channel that always produces the state $\rho$. Hence, this W is Markovian with respect to a DAG with arrows representing such channels, from ${\Gamma^M}$ to ${M_I}$, but is also Markovian to a DAG without such arrows.

If every arrow in the DAG corresponds to a non-trivial channel in the process matrix, the DAG is called \emph{minimal}. From another perspective, a DAG is minimal if, by removing any arrow from it, then the $W$ is not any more Markov with respect to the resulting DAG.

The fact that the output of the code is always the minimal DAG is guaranteed by the first step of the algorithm, where the open subsystems are established and discarded. Indeed, an ``extra arrow'' in a non-minimal DAG would necessarily be associated with an open subsystem---an identity tensor factor in the process matrix.

Note also that, in \cite{costa2016}, it was proven that a DAG can be in principle recovered under the additional assumption of \emph{faithfulness}. Our algorithm does not require such an extra assumption, proving that causal discovery is always possible for a quantum Markov causal model.

\section{Complexity of the algorithm}
The dimension of the process matrix is given by the product of input and output dimension of each party. Thus, the size of the process matrix would generally scale exponentially with the number of parties. This is expected, as also the dimension of ordinary density matrices would scale exponentially with the number of parties. 

One can however consider situations where, under appropriate assumptions and approximations, the physical scenario under consideration is described by a polynomial number of parameters. Then, the main cost of the algorithm lies in two parts: the one that establishes the non-signaling sets and the one that searches for causal arrows between parties. The first step tests condition \eqref{eq:con1} for all parties, to determine each non-signaling set, and the second step tests condition \eqref{eq:con2} (or \eqref{eq:con2i}) for pairs of nodes---in both cases the number of tests required is thus quadratic in the number of parties. Therefore, given an efficient encoding of the input process matrix, the algorithm scales quadratically with the number of parties.

\section{non-Markovian processes}

A Markovian process is one with a process matrix of the form of Equation~\eqref{eq:markovW}, and is represented by a DAG. In a non-Markovian process the process matrix is not of that form, i.e.\ it is not a tensor product of factors representing input states for the parties with no incoming arrows, channels, and identity matrices for the output of the parties in the last set. In other words, in a non-Markovian process, these factors alone---or their representation in a DAG---cannot account for the observed correlations between the events. 

\subsection{Latent nodes}
{If the code outputs that the process is causally ordered but non-Markovian, it can be represented as a quantum circuit compatible with the causal order, where the parties are connected with quantum channels with memory~\cite{Kretschmann2005, chiribella09b}, as we depict in Figure~\ref{fig:circlat} ($\alpha$). {In the language of causal modelling, such a process can be represented by an extended DAG with}
%However, this process can always be extended with 
additional nodes, {called \emph{latent}}, and channels connecting them to the rest of the parties{, so that the extended process is Markovian and reduces to the original one for a particular choice of CPTP maps applied in the extra nodes~\cite{costa2016}, as depicted in } 
%in a way that the whole process is Markovian (maybe a reference? \comment{the reference is our causal modelling paper, but it's already referenced below}) which we depict in 
Figure~\ref{fig:circlat} ($\beta$). The intuition %behind it 
is that the correlations obtained from the original process cannot be produced by considering the original %amount of 
nodes and channels without memory. Therefore, there are extra nodes, not considered in the process, which affect the local outcomes of the nodes considered.} %These are called latent nodes

\bef
\includegraphics[width=\linewidth]{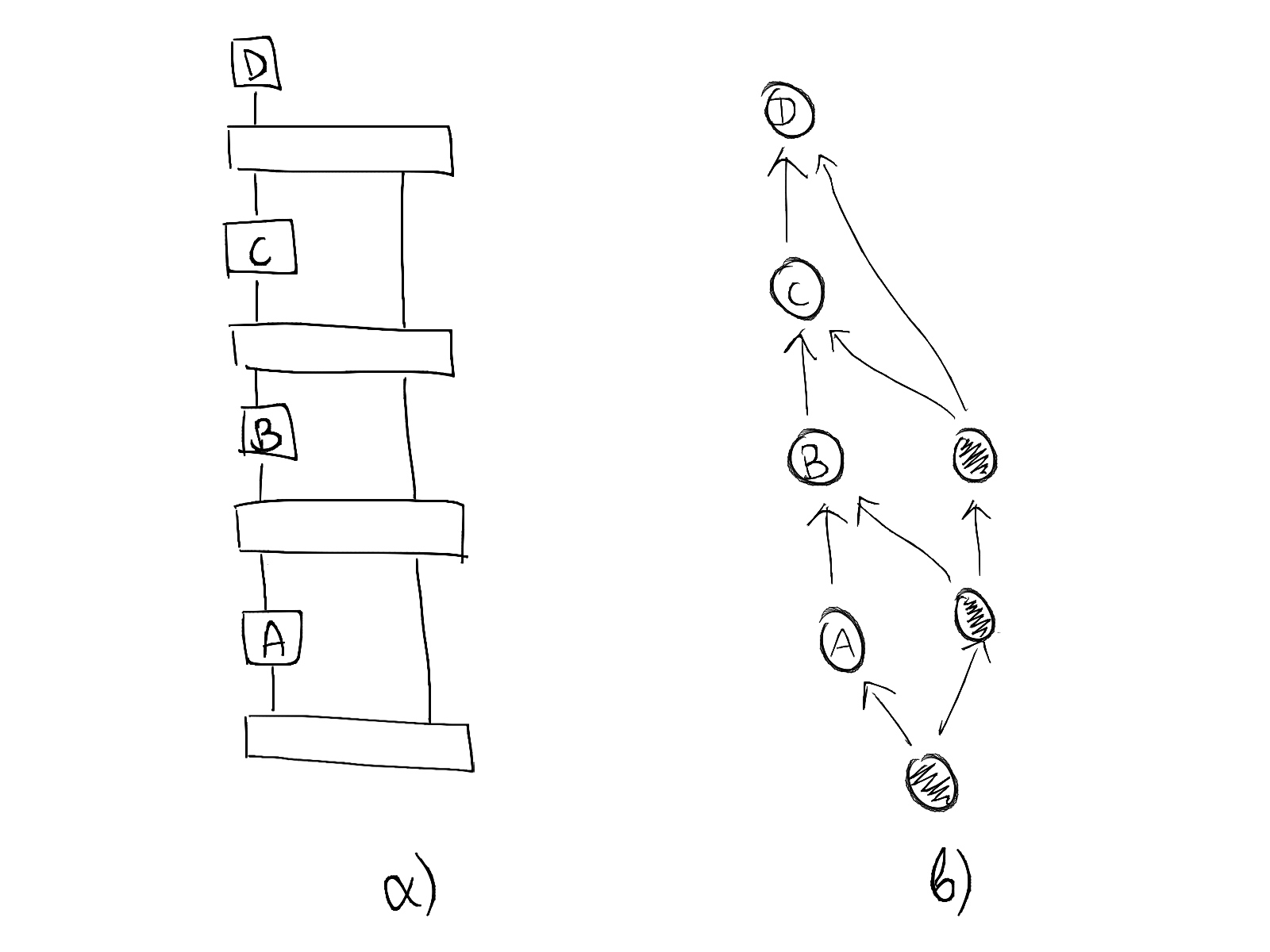}
\caption{{\bf Non-Markovian vs Markovian process: }In figure ($\alpha$), we represent a causally ordered non-Markovian process as a quantum circuit where the channels connecting the parties are quantum channels with memory. In figure ($\beta$), the same process can be represented as a Markovian process for the extended number of nodes. The new nodes introduced are the latent nodes.}
   \label{fig:circlat}
\eef
% 
%\bef
%\includegraphics[width=.9\linewidth]{circlat.jpeg}
%\caption[]{{\bf Non-Markovian vs Markovian process: }In figure $\alpha$), we represent a causally ordered non-Markovian process as a quantum circuit where the channels connecting the parties are quantum channels with memory. In figure $\beta$), the same process can be represented as a Markovian process for the extended number of nodes. The new nodes introduced are the latent nodes. Credits: C. Giarmatzi.}
%\label{fig:circlat}
%\eef

For example, the outcomes of quantum measurements performed in some measurement stations (nodes) in a laboratory, may be affected by the temperature or maybe another system is leaking into one of the stations, like stray light affecting the detection part and causing correlated noise. If these are producing significant change in the data---higher than the noise tolerance in the code---the process will appear non-Markovian.

%Note that the causal arrows found by the causal discovery algorithm, for the considered nodes, may be correct but cannot be accounted for the exact correlations encoded in the process matrix.

To recover a causal model by introducing latent nodes we would need to extend the algorithm such that it adds nodes and arrows until it finds that is Markovian. Computationally, this task can be hard %is very difficult 
because the code has to find the right combination of the number of nodes needed, their position in the DAG and the exact channels around them. However, although the original process is non-Markovian, the code still outputs the causal order of the parties for a causally ordered process matrix. From that, one could make guesses about the right causal model, by introducing nodes with specific input and output systems and channels connecting them to the rest of the parties. To do this, one should add the corresponding factors into the current test matrix $W_{\text{test}}$ and run the code using as input the updated number of parties, dimensions of systems and $W_{\text{test}}$ as the process matrix and see if now the process is Markovian.

\subsection{Mixture of causal orders}
Another possible reason why the process is non-Markovian is that it might be the case that the it represents a probabilistic mixture of two or more Markovian processes with different causal orders, resulting in a non-causally ordered process matrix\footnote{A mixture of processes with the same causal order can be modeled as a causally ordered, non-Markovian process with latent nodes acting as ``classical common causes'' \cite{MacLean2016, Feix2016}}. There is a Semidefinite Program (SDP) for this problem, that finds the right decomposition \cite{araujo15}. For instance, for a bipartite process, the SDP would look like the following.
\ba
\texttt{given}~~&&W \nonumber \\
\texttt{find}~~&&q\\
\texttt{such that}~~&&W = q W^{A\prec B} + (1-q) W^{B\prec A} \nonumber \\
&&0 \leq q \leq 1 \nonumber
\ea
where $W^{X\prec Y}$ denotes a valid process matrix where $Y$ is last and therefore has a factor $\id^{Y_O}$. In the case with more parties, one simply has to write a decomposition that includes all different causal orders for the given parties. Given the result, one can apply the causal discovery algorithm to each term in the decomposition.

\subsection{Dynamical and indefinite causal order}
So far, we have seen that when events have a definite causal order, they can be represented either by a fixed causal order process or by a mixture of causal orders. However,  it may be the case that the process matrix represents a situation of more than two parties, where the causal order of some parties depend on the operations of parties in their past. That is, a party may influence the causal order of future parties. Such a dynamical causal order was studied in~\cite{Oreshkov16} where a definition of causality was proposed, compatible with such dynamical causal order. For the tripartite case, it was found that the process matrix describing such a situation should obey certain conditions. However similar conditions were not found for the case of arbitrary parties. In such cases, the notion of causal discovery is not clear, as depending on some events in the past, the DAG of future ones would change. Hence the output would be different DAGs for different operations of certain parties. We do not know if the discovery of those DAGs is possible.

\subsection{Different definitions of Markovianity}
Our algorithm relies on the definition of quantum Markov causal model of Ref.~\cite{costa2016}. A different definition was proposed in Ref.~\cite{Allen2016}, where the output systems of the parties are not assumed to factorize into subsystems in the presence of multiple outgoing arrows. In Ref.~\cite{Allen2016}, arrows in the DAG are still associated with a quantum channel from the output space of the parent nodes to the input space of the child but, rather than defining a factorization in subsystems of the output space, multiple outgoing arrows are more generally associated with commuting channels. For example, in a tripartite scenario where $A$ is a parent of both $B$ and $C$, a Markovian process matrix would have the form 
\begin{multline}
W^{A_IA_OB_IB_OC_IC_O} \\
= \rho^{A_I}\otimes\left(T_1^{A_OB_I} \cdot T_2^{A_OC_I}\right)\otimes\id^{B_OC_O},
\label{allenmarkov}
\end{multline}
with the condition $T_1^{A_OB_I}\cdot T_2^{A_OC_I} = T_2^{A_OC_I}\cdot T_1^{A_OB_I}$. Thus, according to Ref.~\cite{Allen2016}, a Markovian process matrix does not need to be a tensor product but can more generally be a product of commuting matrices. To distinguish the two definitions, we will call \emph{tensor-Markovian}  and  \emph{commuting-Markovian} a process matrix that satisfies the condition of Ref.~\cite{costa2016} (used in our code) and Ref.~\cite{Allen2016}, respectively. Note that all tensor-Markovian processes are commuting Markovian, but the converse is not true\footnote{In Ref.~\cite{Allen2016} it is further assumed that input and output spaces of each node are isomorphic. Thus, strictly speaking, not all tensor-Markovian process considered here satisfy the definition of Ref.~\cite{Allen2016}, but only those with input and output of equal dimension. This difference is of little consequence from the point of view of a causal discovery algorithm, since in any case the dimension of each space has to be specified as input to the code.}.

Our algorithm could be adapted to discover the causal structure of commuting-Markovian processes. Note that the strategy used in our code, to detect the parent space of each node by checking~\eqref{eq:channel}, would not work. Indeed, tracing out $B_I$ from matrix \eqref{allenmarkov} does not result in a matrix with identity on $A_O$. A possible approach could be to instead detect all the \emph{children} of each node $A$, namely all the nodes with an incoming arrow departing from $A$. The children are then identified as the smallest subset of parties $C^1,\dots,C^k$ such that 
\begin{equation}
\tilde{\id}^{A_{O}} \otimes \Tr_{A_{O}}\left( \Tr_{C_I^1,\dots,C_I^k}W\right) =  \Tr_{C_I^1,\dots,C_I^k}W.
\label{children}
\end{equation}
As this condition must be checked for subsets of parties, the number of tests is exponential in the number of parties for the worst-case scenario. In contrast, we have seen that to discover a tensor-Markovian causal structure a quadratic number of tests is sufficient. Another potential complication is that our test for Markovianity relies on the tensor-product form of the process matrices; it is not clear if there is a simple way to test whether a process is commuting-Markovian. 

An alternative approach is to retain the definition of tensor-Markovian processes and model commuting-Markovian processes as non-Markovian ones. Indeed, since a commuting-Markovian process is causally ordered, it can always be recovered from a tensor-Markovian one by adding an appropriate number of latent nodes~\cite{costa2016}. An extension of our code to detect latent nodes could thus be used to detect the causal structure of a commuting-Markovian process. In figure~\ref{fig:latent} we show an example of a DAG of a commuting-Markovian process (left) and how that would be represented as a tensor-Markovian (right) with a latent node.

\bef
\includegraphics[width=\linewidth]{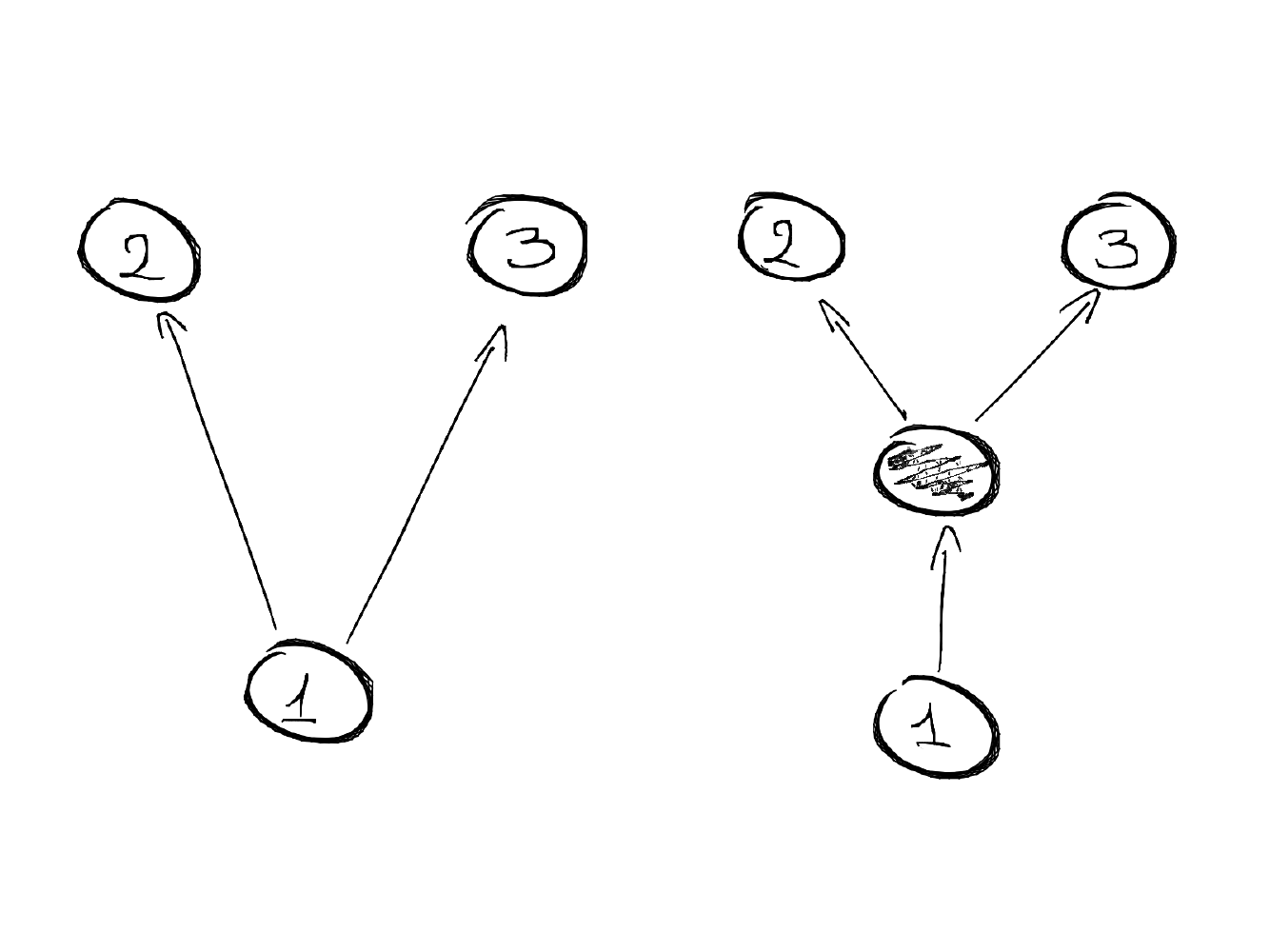}
\caption[]{\textbf{Different definitions of Markovianity}: A process that is Markovian according to Ref.~\cite{Allen2016}, e.g.\ for the DAG on the left, is generally described by a DAG with a latent node (filled node in the DAG on the right) according to the definition of Markovianity of Ref.~\cite{costa2016} on which our algorithm is based.}
\label{fig:latent}
\eef

\section{Conclusions}
We have presented an algorithm (whose implementation we provide) that can discover an initially unknown causal structure in a quantum network. The first of its type, it is an important proof of principle: it shows that causal structure has a precise empirical meaning in quantum mechanics. Just as other physical properties, it can be unknown and discovered. This is of particular significance for foundational approaches where causal structure is seen as emergent from more fundamental primitives. Causal discovery provides the methodology to determine when and how causal structure emerges.

Causal discovery can also have broad applications for protocols based on large and complex quantum networks. Our algorithm is guaranteed to find a minimal causal model for any Markovian process, namely a process in which all causally relevant events are under experimental control,  with no extra assumptions; this improves on the results of Ref.~\cite{costa2016}, where the additional condition of faithfulness was invoked. Even for non-Markovian processes, the algorithm still recovers important causal information, namely a causal order of the events.

Another important use of our algorithm is to tackle the difficult problem of non-Markovianity. An extensive body of research is currently devoted to the problem of detecting non-Markovianity~\cite{Rivas2014}. Our algorithm finds a concrete solution: it allows discovering when some external memory is affecting the correlations in the observed system. Detecting non-Markovianity can also have important practical applications for large quantum networks: the presence of ``latent nodes'', can indicate a possible source of systematic correlated noise in a process, that might affect the working of a quantum protocol. It can further have applications in cryptography for detecting the presence of an eavesdropper.

Finally, our algorithm has promising possible extensions. A natural extension is an algorithm that can make ``good guesses'' for causal structure in the presence of latent nodes. Promising is also the extension of causal discovery to mixtures of causal order, dynamical and indefinite causal structure. 
\\
\begin{acknowledgments}
We thank Gerard Milburn, Sally Shrapnel and Andrew White for discussions. This work was supported by the Australian Research Council (ARC) Centre of Excellence for Quantum Engineered Systems grant (CE 110001013), the ARC Centre for Quantum Computation and Communication Technology (Grant No. CE110001027) and by the Templeton World Charity Foundation (TWCF 0064/AB38). Furthermore, this publication was made possible through the support of a grant from the John Templeton Foundation. The opinions expressed in this publication are those of the authors and do not necessarily reflect the views of the John Templeton Foundation. We acknowledge the traditional owners of the land on which the University of Queensland is situated, the Turrbal and Jagera people.
\end{acknowledgments}

%\bibliography{DiscoveryBib}

\begin{thebibliography}{34}%
\makeatletter
\providecommand \@ifxundefined [1]{%
 \@ifx{#1\undefined}
}%
\providecommand \@ifnum [1]{%
 \ifnum #1\expandafter \@firstoftwo
 \else \expandafter \@secondoftwo
 \fi
}%
\providecommand \@ifx [1]{%
 \ifx #1\expandafter \@firstoftwo
 \else \expandafter \@secondoftwo
 \fi
}%
\providecommand \natexlab [1]{#1}%
\providecommand \enquote  [1]{``#1''}%
\providecommand \bibnamefont  [1]{#1}%
\providecommand \bibfnamefont [1]{#1}%
\providecommand \citenamefont [1]{#1}%
\providecommand \href@noop [0]{\@secondoftwo}%
\providecommand \href [0]{\begingroup \@sanitize@url \@href}%
\providecommand \@href[1]{\@@startlink{#1}\@@href}%
\providecommand \@@href[1]{\endgroup#1\@@endlink}%
\providecommand \@sanitize@url [0]{\catcode `\\12\catcode `\$12\catcode
  `\&12\catcode `\#12\catcode `\^12\catcode `\_12\catcode `\%12\relax}%
\providecommand \@@startlink[1]{}%
\providecommand \@@endlink[0]{}%
\providecommand \url  [0]{\begingroup\@sanitize@url \@url }%
\providecommand \@url [1]{\endgroup\@href {#1}{\urlprefix }}%
\providecommand \urlprefix  [0]{URL }%
\providecommand \Eprint [0]{\href }%
\providecommand \doibase [0]{http://dx.doi.org/}%
\providecommand \selectlanguage [0]{\@gobble}%
\providecommand \bibinfo  [0]{\@secondoftwo}%
\providecommand \bibfield  [0]{\@secondoftwo}%
\providecommand \translation [1]{[#1]}%
\providecommand \BibitemOpen [0]{}%
\providecommand \bibitemStop [0]{}%
\providecommand \bibitemNoStop [0]{.\EOS\space}%
\providecommand \EOS [0]{\spacefactor3000\relax}%
\providecommand \BibitemShut  [1]{\csname bibitem#1\endcsname}%
\let\auto@bib@innerbib\@empty
%</preamble>
\bibitem [{\citenamefont {Pearl}(2009)}]{Pearlbook}%
  \BibitemOpen
  \bibfield  {author} {\bibinfo {author} {\bibfnamefont {J.}~\bibnamefont
  {Pearl}},\ }\href@noop {} {\emph {\bibinfo {title} {Causality.}}}\ (\bibinfo
  {publisher} {Cambridge University Press},\ \bibinfo {year}
  {2009})\BibitemShut {NoStop}%
\bibitem [{\citenamefont {Spirtes}\ \emph {et~al.}(2000)\citenamefont
  {Spirtes}, \citenamefont {Glymour},\ and\ \citenamefont
  {Scheines}}]{spirtes2000causation}%
  \BibitemOpen
  \bibfield  {author} {\bibinfo {author} {\bibfnamefont {P.}~\bibnamefont
  {Spirtes}}, \bibinfo {author} {\bibfnamefont {C.~N.}\ \bibnamefont
  {Glymour}}, \ and\ \bibinfo {author} {\bibfnamefont {R.}~\bibnamefont
  {Scheines}},\ }\href@noop {} {\emph {\bibinfo {title} {Causation, prediction,
  and search}}},\ Vol.~\bibinfo {volume} {81}\ (\bibinfo  {publisher} {MIT
  press},\ \bibinfo {year} {2000})\BibitemShut {NoStop}%
\bibitem [{\citenamefont {Lamport}(1978)}]{Lamport1978}%
  \BibitemOpen
  \bibfield  {author} {\bibinfo {author} {\bibfnamefont {L.}~\bibnamefont
  {Lamport}},\ }\bibfield  {title} {\enquote {\bibinfo {title} {Time, clocks,
  and the ordering of events in a distributed system},}\ }\href {\doibase
  10.1145/359545.359563} {\bibfield  {journal} {\bibinfo  {journal} {Commun.
  ACM}\ }\textbf {\bibinfo {volume} {21}},\ \bibinfo {pages} {558--565}
  (\bibinfo {year} {1978})}\BibitemShut {NoStop}%
\bibitem [{\citenamefont {Wood}\ and\ \citenamefont
  {Spekkens}(2015)}]{woodlesson2012}%
  \BibitemOpen
  \bibfield  {author} {\bibinfo {author} {\bibfnamefont {C.~J.}\ \bibnamefont
  {Wood}}\ and\ \bibinfo {author} {\bibfnamefont {R.~W.}\ \bibnamefont
  {Spekkens}},\ }\bibfield  {title} {\enquote {\bibinfo {title} {The lesson of
  causal discovery algorithms for quantum correlations: {Causal} explanations
  of {Bell}-inequality violations require fine-tuning},}\ }\href {\doibase
  10.1088/1367-2630/17/3/033002} {\bibfield  {journal} {\bibinfo  {journal}
  {New J. Phys.}\ }\textbf {\bibinfo {volume} {17}},\ \bibinfo {pages} {033002}
  (\bibinfo {year} {2015})}\BibitemShut {NoStop}%
\bibitem [{\citenamefont {Tucci}(1995)}]{tucciquantum1995}%
  \BibitemOpen
  \bibfield  {author} {\bibinfo {author} {\bibfnamefont {R.~R.}\ \bibnamefont
  {Tucci}},\ }\bibfield  {title} {\enquote {\bibinfo {title} {Quantum
  {Bayesian} {Nets}},}\ }\href {\doibase 10.1142/S0217979295000148} {\bibfield
  {journal} {\bibinfo  {journal} {Int. J. Mod. Phys. B}\ }\textbf {\bibinfo
  {volume} {09}},\ \bibinfo {pages} {295--337} (\bibinfo {year} {1995})},\
  \bibinfo {note} {arXiv: quant-ph/9706039}\BibitemShut {NoStop}%
\bibitem [{\citenamefont {Leifer}(2006)}]{Leifer2006}%
  \BibitemOpen
  \bibfield  {author} {\bibinfo {author} {\bibfnamefont {M.~S.}\ \bibnamefont
  {Leifer}},\ }\bibfield  {title} {\enquote {\bibinfo {title} {Quantum dynamics
  as an analog of conditional probability},}\ }\href {\doibase
  10.1103/PhysRevA.74.042310} {\bibfield  {journal} {\bibinfo  {journal} {Phys.
  Rev. A}\ }\textbf {\bibinfo {volume} {74}},\ \bibinfo {pages} {042310}
  (\bibinfo {year} {2006})}\BibitemShut {NoStop}%
\bibitem [{\citenamefont {{Laskey}}(2007)}]{Laskey2007}%
  \BibitemOpen
  \bibfield  {author} {\bibinfo {author} {\bibfnamefont {K.~B.}\ \bibnamefont
  {{Laskey}}},\ }\bibfield  {title} {\enquote {\bibinfo {title} {{Quantum
  Causal Networks}},}\ }\href@noop {} {\  (\bibinfo {year} {2007})}, Preprint at \Eprint
  {http://arxiv.org/abs/0710.1200} {arXiv:0710.1200 [quant-ph]} \BibitemShut
  {NoStop}%
\bibitem [{\citenamefont {Leifer}\ and\ \citenamefont
  {Spekkens}(2013)}]{Leifer2013}%
  \BibitemOpen
  \bibfield  {author} {\bibinfo {author} {\bibfnamefont {M.~S.}\ \bibnamefont
  {Leifer}}\ and\ \bibinfo {author} {\bibfnamefont {R.~W.}\ \bibnamefont
  {Spekkens}},\ }\bibfield  {title} {\enquote {\bibinfo {title} {Towards a
  formulation of quantum theory as a causally neutral theory of bayesian
  inference},}\ }\href {\doibase 10.1103/PhysRevA.88.052130} {\bibfield
  {journal} {\bibinfo  {journal} {Physical Review A}\ }\textbf {\bibinfo
  {volume} {88}},\ \bibinfo {pages} {052130} (\bibinfo {year}
  {2013})}\BibitemShut {NoStop}%
\bibitem [{\citenamefont {Cavalcanti}\ and\ \citenamefont
  {Lal}(2014)}]{cavalcanti2014modifications}%
  \BibitemOpen
  \bibfield  {author} {\bibinfo {author} {\bibfnamefont {E.~G.}\ \bibnamefont
  {Cavalcanti}}\ and\ \bibinfo {author} {\bibfnamefont {R.}~\bibnamefont
  {Lal}},\ }\bibfield  {title} {\enquote {\bibinfo {title} {On modifications of
  reichenbach's principle of common cause in light of bell's theorem.}}\ }\href
  {\doibase 10.1088/1751-8113/47/42/424018} {\bibfield  {journal} {\bibinfo
  {journal} {J.\ Phys.\ A: Math.\ Theor.}\ }\textbf {\bibinfo {volume} {47}},\
  \bibinfo {pages} {424018} (\bibinfo {year} {2014})}\BibitemShut {NoStop}%
\bibitem [{\citenamefont {Fritz}(2015)}]{fritzbeyond2015}%
  \BibitemOpen
  \bibfield  {author} {\bibinfo {author} {\bibfnamefont {T.}~\bibnamefont
  {Fritz}},\ }\bibfield  {title} {\enquote {\bibinfo {title} {Beyond bell's
  theorem {II}: Scenarios with arbitrary causal structure},}\ }\href {\doibase
  10.1007/s00220-015-2495-5} {\bibfield  {journal} {\bibinfo  {journal} {Comm.
  Math. Phys.}\ ,\ \bibinfo {pages} {1--44}} (\bibinfo {year}
  {2015})}\BibitemShut {NoStop}%
\bibitem [{\citenamefont {Henson}\ \emph {et~al.}(2014)\citenamefont {Henson},
  \citenamefont {Lal},\ and\ \citenamefont {Pusey}}]{Henson2015}%
  \BibitemOpen
  \bibfield  {author} {\bibinfo {author} {\bibfnamefont {J.}~\bibnamefont
  {Henson}}, \bibinfo {author} {\bibfnamefont {R.}~\bibnamefont {Lal}}, \ and\
  \bibinfo {author} {\bibfnamefont {M.~F}\ \bibnamefont {Pusey}},\ }\bibfield
  {title} {\enquote {\bibinfo {title} {Theory-independent limits on
  correlations from generalized bayesian networks.}}\ }\href {\doibase
  10.1088/1367-2630/16/11/113043} {\bibfield  {journal} {\bibinfo  {journal}
  {New\ J.\ Phys.}\ }\textbf {\bibinfo {volume} {16}},\ \bibinfo {pages}
  {113043} (\bibinfo {year} {2014})}\BibitemShut {NoStop}%
\bibitem [{\citenamefont {Pienaar}\ and\ \citenamefont
  {Brukner}(2015)}]{pienaar2014graph}%
  \BibitemOpen
  \bibfield  {author} {\bibinfo {author} {\bibfnamefont {J.}~\bibnamefont
  {Pienaar}}\ and\ \bibinfo {author} {\bibfnamefont {{\v C}.}~\bibnamefont
  {Brukner}},\ }\bibfield  {title} {\enquote {\bibinfo {title} {A
  graph-separation theorem for quantum causal models.}}\ }\href {\doibase
  10.1088/1367-2630/17/7/073020} {\bibfield  {journal} {\bibinfo  {journal}
  {New\ J.\ Phys.}\ }\textbf {\bibinfo {volume} {17}},\ \bibinfo {pages}
  {073020} (\bibinfo {year} {2015})}\BibitemShut {NoStop}%
\bibitem [{\citenamefont {Chaves}\ \emph {et~al.}(2015)\citenamefont {Chaves},
  \citenamefont {Majenz},\ and\ \citenamefont {Gross}}]{chavesinformation2015}%
  \BibitemOpen
  \bibfield  {author} {\bibinfo {author} {\bibfnamefont {R.}~\bibnamefont
  {Chaves}}, \bibinfo {author} {\bibfnamefont {C.}~\bibnamefont {Majenz}}, \
  and\ \bibinfo {author} {\bibfnamefont {D.}~\bibnamefont {Gross}},\ }\bibfield
   {title} {{\selectlanguage {UKenglish}\enquote {\bibinfo {title}
  {Information--theoretic implications of quantum causal structures},}\ }}\href
  {\doibase 10.1038/ncomms6766} {\bibfield  {journal} {\bibinfo  {journal}
  {Nat. Commun.}\ }\textbf {\bibinfo {volume} {6}} (\bibinfo {year} {2015}),\
  10.1038/ncomms6766}\BibitemShut {NoStop}%
\bibitem [{\citenamefont {Ried}\ \emph {et~al.}(2015)\citenamefont {Ried},
  \citenamefont {Agnew}, \citenamefont {Vermeyden}, \citenamefont {Janzing},
  \citenamefont {Spekkens},\ and\ \citenamefont {Resch}}]{ried2015quantum}%
  \BibitemOpen
  \bibfield  {author} {\bibinfo {author} {\bibfnamefont {K.}~\bibnamefont
  {Ried}}, \bibinfo {author} {\bibfnamefont {M.}~\bibnamefont {Agnew}},
  \bibinfo {author} {\bibfnamefont {L.}~\bibnamefont {Vermeyden}}, \bibinfo
  {author} {\bibfnamefont {D.}~\bibnamefont {Janzing}}, \bibinfo {author}
  {\bibfnamefont {R.~W.}\ \bibnamefont {Spekkens}}, \ and\ \bibinfo {author}
  {\bibfnamefont {K.~J.}\ \bibnamefont {Resch}},\ }\bibfield  {title} {\enquote
  {\bibinfo {title} {A quantum advantage for inferring causal structure},}\
  }\href {\doibase 10.1038/nphys3266} {\bibfield  {journal} {\bibinfo
  {journal} {Nat. Phys.}\ }\textbf {\bibinfo {volume} {11}},\ \bibinfo {pages}
  {414--420} (\bibinfo {year} {2015})}\BibitemShut {NoStop}%
\bibitem [{\citenamefont {Costa}\ and\ \citenamefont
  {Shrapnel}(2016)}]{costa2016}%
  \BibitemOpen
  \bibfield  {author} {\bibinfo {author} {\bibfnamefont {F.}~\bibnamefont
  {Costa}}\ and\ \bibinfo {author} {\bibfnamefont {S.}~\bibnamefont
  {Shrapnel}},\ }\bibfield  {title} {\enquote {\bibinfo {title} {Quantum causal
  modelling},}\ }\href {\doibase 10.1088/1367-2630/18/6/063032} {\bibfield
  {journal} {\bibinfo  {journal} {New Journal of Physics}\ }\textbf {\bibinfo
  {volume} {18}},\ \bibinfo {pages} {063032} (\bibinfo {year}
  {2016})}\BibitemShut {NoStop}%
\bibitem{Allen2016}
J.-M.~A. Allen, J.~Barrett, D.~C. Horsman, C.~M. Lee, and R.~W. Spekkens.
\newblock Quantum common causes and quantum causal models.
\newblock {\em Physical Review X}, 7(3):031021, 07 2017
  \BibitemShut {NoStop}%
\bibitem [{\citenamefont {Shrapnel}(2015)}]{Shrapnel2015}%
  \BibitemOpen
  \bibfield  {author} {\bibinfo {author} {\bibfnamefont {S.}~\bibnamefont
  {Shrapnel}},\ }\href {http://philsci-archive.pitt.edu/11751/} {\enquote
  {\bibinfo {title} {Discovering quantum causal models},}\ } (\bibinfo {year}
  {2015})\BibitemShut {NoStop}%
\bibitem [{\citenamefont {Shrapnel}(2016)}]{Shrapnel2016}%
  \BibitemOpen
  \bibfield  {author} {\bibinfo {author} {\bibfnamefont {S.}~\bibnamefont
  {Shrapnel}},\ }{\selectlanguage {UKenglish}\emph {\bibinfo {title} {Using
  interventions to discover quantum causal structure}}},\ \href {\doibase
  10.14264/uql.2016.1102} {Ph.D. thesis} (\bibinfo {year} {2016})\BibitemShut
  {NoStop}%
\bibitem [{\citenamefont {{Oreshkov}}\ \emph {et~al.}(2012)\citenamefont
  {{Oreshkov}}, \citenamefont {{Costa}},\ and\ \citenamefont
  {{Brukner}}}]{oreshkov12}%
  \BibitemOpen
  \bibfield  {author} {\bibinfo {author} {\bibfnamefont {O.}~\bibnamefont
  {{Oreshkov}}}, \bibinfo {author} {\bibfnamefont {F.}~\bibnamefont {{Costa}}},
  \ and\ \bibinfo {author} {\bibfnamefont {{\v C}.}~\bibnamefont {{Brukner}}},\
  }\bibfield  {title} {\enquote {\bibinfo {title} {{Quantum correlations with
  no causal order}},}\ }\href {\doibase 10.1038/ncomms2076} {\bibfield
  {journal} {\bibinfo  {journal} {Nat. Commun.}\ }\textbf {\bibinfo {volume}
  {3}},\ \bibinfo {pages} {1092} (\bibinfo {year} {2012})}\BibitemShut
  {NoStop}%
\bibitem [{dis()}]{discovery_code2017}%
  \BibitemOpen
  \href {https://github.com/Christina-Giar/quantum-causal-discovery-algo.git} {\bibinfo  {journal}
  {https://github.com/Christina-Giar/quantum-causal-discovery-algo.git}\ }\BibitemShut {NoStop}%
\bibitem [{Cub()}]{Cubitt}%
  \BibitemOpen
\bibfield  {journal} {  }\href {http://www.dr-qubit.org/Matlab_code} {\bibinfo
  {journal} {http://www.dr-qubit.org/Matlab\_code}\ }\BibitemShut {NoStop}%
\bibitem [{\citenamefont {{Chiribella}}\ \emph {et~al.}(2009)\citenamefont
  {{Chiribella}}, \citenamefont {{D'Ariano}},\ and\ \citenamefont
  {{Perinotti}}}]{chiribella09b}%
  \BibitemOpen
\bibfield  {journal} {  }\bibfield  {author} {\bibinfo {author} {\bibfnamefont
  {G.}~\bibnamefont {{Chiribella}}}, \bibinfo {author} {\bibfnamefont {G.~M.}\
  \bibnamefont {{D'Ariano}}}, \ and\ \bibinfo {author} {\bibfnamefont
  {P.}~\bibnamefont {{Perinotti}}},\ }\bibfield  {title} {\enquote {\bibinfo
  {title} {{Theoretical framework for quantum networks}},}\ }\href {\doibase
  10.1103/PhysRevA.80.022339} {\bibfield  {journal} {\bibinfo  {journal} {Phys.
  Rev.~A}\ }\textbf {\bibinfo {volume} {80}},\ \bibinfo {eid} {022339}
  (\bibinfo {year} {2009})}\BibitemShut {NoStop}%
\bibitem [{\citenamefont {Jamio{\l}kowski}(1972)}]{jamio72}%
  \BibitemOpen
  \bibfield  {author} {\bibinfo {author} {\bibfnamefont {A.}~\bibnamefont
  {Jamio{\l}kowski}},\ }\bibfield  {title} {\enquote {\bibinfo {title} {Linear
  transformations which preserve trace and positive semidefiniteness of
  operators},}\ }\href {\doibase 10.1016/0034-4877(72)90011-0} {\bibfield
  {journal} {\bibinfo  {journal} {Rep. Math. Phys}\ }\textbf {\bibinfo {volume}
  {3}},\ \bibinfo {pages} {275--278} (\bibinfo {year} {1972})}\BibitemShut
  {NoStop}%
\bibitem [{\citenamefont {Choi}(1975)}]{Choi1975}%
  \BibitemOpen
  \bibfield  {author} {\bibinfo {author} {\bibfnamefont {Man-Duen}\
  \bibnamefont {Choi}},\ }\bibfield  {title} {{\selectlanguage
  {UKenglish}\enquote {\bibinfo {title} {Completely positive linear maps on
  complex matrices},}\ }}\href {\doibase 10.1016/0024-3795(75)90075-0}
  {\bibfield  {journal} {\bibinfo  {journal} {Linear Algebra Appl.}\ }\textbf
  {\bibinfo {volume} {10}},\ \bibinfo {pages} {285--290} (\bibinfo {year}
  {1975})}\BibitemShut {NoStop}%
\bibitem [{\citenamefont {{Caves}}\ \emph {et~al.}(2004)\citenamefont
  {{Caves}}, \citenamefont {{Fuchs}}, \citenamefont {{Manne}},\ and\
  \citenamefont {{Renes}}}]{caves2004}%
  \BibitemOpen
  \bibfield  {author} {\bibinfo {author} {\bibfnamefont {C.~M.}\ \bibnamefont
  {{Caves}}}, \bibinfo {author} {\bibfnamefont {C.~A.}\ \bibnamefont
  {{Fuchs}}}, \bibinfo {author} {\bibfnamefont {K.~K.}\ \bibnamefont
  {{Manne}}}, \ and\ \bibinfo {author} {\bibfnamefont {J.~M.}\ \bibnamefont
  {{Renes}}},\ }\bibfield  {title} {\enquote {\bibinfo {title} {{Gleason-Type
  Derivations of the Quantum Probability Rule for Generalized Measurements}},}\
  }\href {\doibase 10.1023/B:FOOP.0000019581.00318.a5} {\bibfield  {journal}
  {\bibinfo  {journal} {Found. Phys.}\ }\textbf {\bibinfo {volume} {34}},\
  \bibinfo {pages} {193--209} (\bibinfo {year} {2004})}\BibitemShut {NoStop}%
\bibitem [{\citenamefont {Shrapnel}\ \emph {et~al.}(2017)\citenamefont
  {Shrapnel}, \citenamefont {Costa},\ and\ \citenamefont
  {Milburn}}]{shrapnel2017updating}%
  \BibitemOpen
  \bibfield  {author} {\bibinfo {author} {\bibfnamefont {S.}\ \bibnamefont
  {Shrapnel}}, \bibinfo {author} {\bibfnamefont {F.}\ \bibnamefont {Costa}},
  \ and\ \bibinfo {author} {\bibfnamefont {G.}\ \bibnamefont {Milburn}},\
  }\bibfield  {title} {\enquote {\bibinfo {title} {Updating the {Born} rule},}\
  }\href@noop {} {\  (\bibinfo {year} {2017})},\ Preprint at \Eprint
  {http://arxiv.org/abs/1702.01845} {arXiv:1702.01845 [quant-ph]} \BibitemShut
  {NoStop}%
\bibitem [{\citenamefont {Pollock}\ \emph {et~al.}(2015)\citenamefont
  {Pollock}, \citenamefont {Rodr{\'\i}guez-Rosario}, \citenamefont
  {Frauenheim}, \citenamefont {Paternostro},\ and\ \citenamefont
  {Modi}}]{pollockcomplete2015}%
  \BibitemOpen
  \bibfield  {author} {\bibinfo {author} {\bibfnamefont {F.~A.}\
  \bibnamefont {Pollock}}, \bibinfo {author} {\bibfnamefont {C.}\
  \bibnamefont {Rodr{\'\i}guez-Rosario}}, \bibinfo {author} {\bibfnamefont
  {T.}\ \bibnamefont {Frauenheim}}, \bibinfo {author} {\bibfnamefont
  {M.}\ \bibnamefont {Paternostro}}, \ and\ \bibinfo {author} {\bibfnamefont
  {K.}\ \bibnamefont {Modi}},\ }\bibfield  {title} {\enquote {\bibinfo
  {title} {Complete framework for efficient characterisation of non-{Markovian}
  processes},}\ }\href@noop {} {\  (\bibinfo {year} {2015})},\ Preprint at \Eprint
  {http://arxiv.org/abs/1512.00589} {arXiv:1512.00589 [quant-ph]} \BibitemShut
  {NoStop}%
\bibitem [{\citenamefont {Gutoski}\ and\ \citenamefont
  {Watrous}(2006)}]{gutoski06}%
  \BibitemOpen
  \bibfield  {author} {\bibinfo {author} {\bibfnamefont {G.}\ \bibnamefont
  {Gutoski}}\ and\ \bibinfo {author} {\bibfnamefont {J.}\ \bibnamefont
  {Watrous}},\ }\bibfield  {title} {\enquote {\bibinfo {title} {Toward a
  general theory of quantum games},}\ }in\ \href@noop {} {\emph {\bibinfo
  {booktitle} {Proceedings of 39th ACM STOC}}}\ (\bibinfo {year} {2006})\ pp.\
  \bibinfo {pages} {565--574},\ Preprint at \Eprint {http://arxiv.org/abs/quant-ph/0611234}
  {arXiv:quant-ph/0611234} \BibitemShut {NoStop}%
\bibitem [{\citenamefont {Kretschmann}\ and\ \citenamefont
  {Werner}(2005)}]{Kretschmann2005}%
  \BibitemOpen
  \bibfield  {author} {\bibinfo {author} {\bibfnamefont {D.}~\bibnamefont
  {Kretschmann}}\ and\ \bibinfo {author} {\bibfnamefont {R.~F.}\ \bibnamefont
  {Werner}},\ }\bibfield  {title} {\enquote {\bibinfo {title} {Quantum channels
  with memory},}\ }\href {\doibase 10.1103/PhysRevA.72.062323} {\bibfield
  {journal} {\bibinfo  {journal} {Phys. Rev. A}\ }\textbf {\bibinfo {volume}
  {72}},\ \bibinfo {pages} {062323} (\bibinfo {year} {2005})}\BibitemShut
  {NoStop}%
\bibitem [{\citenamefont {Ara{\'u}jo}\ \emph {et~al.}(2015)\citenamefont
  {Ara{\'u}jo}, \citenamefont {Branciard}, \citenamefont {Costa}, \citenamefont
  {Feix}, \citenamefont {Giarmatzi},\ and\ \citenamefont {Brukner}}]{araujo15}%
  \BibitemOpen
  \bibfield  {author} {\bibinfo {author} {\bibfnamefont {M.}~\bibnamefont
  {Ara{\'u}jo}}, \bibinfo {author} {\bibfnamefont {C.}~\bibnamefont
  {Branciard}}, \bibinfo {author} {\bibfnamefont {F.}~\bibnamefont {Costa}},
  \bibinfo {author} {\bibfnamefont {A.}~\bibnamefont {Feix}}, \bibinfo {author}
  {\bibfnamefont {C.}~\bibnamefont {Giarmatzi}}, \ and\ \bibinfo {author}
  {\bibfnamefont {{\v C}.}~\bibnamefont {Brukner}},\ }\bibfield  {title}
  {\enquote {\bibinfo {title} {Witnessing causal nonseparability},}\ }\href
  {\doibase 10.1088/1367-2630/17/10/102001} {\bibfield  {journal} {\bibinfo
  {journal} {New J. Phys.}\ }\textbf {\bibinfo {volume} {17}},\ \bibinfo
  {pages} {102001} (\bibinfo {year} {2015})}\BibitemShut {NoStop}%
\bibitem [{\citenamefont {MacLean}\ \emph {et~al.}(2016)\citenamefont
  {MacLean}, \citenamefont {Ried}, \citenamefont {Spekkens},\ and\
  \citenamefont {Resch}}]{MacLean2016}%
  \BibitemOpen
  \bibfield  {author} {\bibinfo {author} {\bibfnamefont {J.-P.~W.}\
  \bibnamefont {MacLean}}, \bibinfo {author} {\bibfnamefont {K.}~\bibnamefont
  {Ried}}, \bibinfo {author} {\bibfnamefont {R.~W.}\ \bibnamefont {Spekkens}},
  \ and\ \bibinfo {author} {\bibfnamefont {K.~J.}\ \bibnamefont {Resch}},\
  }\bibfield  {title} {\enquote {\bibinfo {title} {Quantum-coherent mixtures of
  causal relations},}\ }\href@noop {} {\  (\bibinfo {year} {2016})},\ Preprint at \Eprint
  {http://arxiv.org/abs/1606.04523} {arXiv:1606.04523 [quant-ph]} \BibitemShut
  {NoStop}%
\bibitem [{\citenamefont {Feix}\ and\ \citenamefont
  {Brukner}(2016)}]{Feix2016}%
  \BibitemOpen
  \bibfield  {author} {\bibinfo {author} {\bibfnamefont {A.}~\bibnamefont
  {Feix}}\ and\ \bibinfo {author} {\bibfnamefont {{\v C}.}~\bibnamefont
  {Brukner}},\ }\bibfield  {title} {\enquote {\bibinfo {title} {Quantum
  superpositions of "common-cause" and "direct-cause" causal structures},}\
  }\href@noop {} {\  (\bibinfo {year} {2016})},\ Preprint at \Eprint
  {http://arxiv.org/abs/1606.09241} {arXiv:1606.09241 [quant-ph]} \BibitemShut
  {NoStop}%
\bibitem [{\citenamefont {Oreshkov}\ and\ \citenamefont
  {Giarmatzi}(2016)}]{Oreshkov16}%
  \BibitemOpen
  \bibfield  {author} {\bibinfo {author} {\bibfnamefont {Ognyan}\ \bibnamefont
  {Oreshkov}}\ and\ \bibinfo {author} {\bibfnamefont {Christina}\ \bibnamefont
  {Giarmatzi}},\ }\bibfield  {title} {\enquote {\bibinfo {title} {Causal and
  causally separable processes},}\ }\href
  {http://stacks.iop.org/1367-2630/18/i=9/a=093020} {\bibfield  {journal}
  {\bibinfo  {journal} {New Journal of Physics}\ }\textbf {\bibinfo {volume}
  {18}},\ \bibinfo {pages} {093020} (\bibinfo {year} {2016})}\BibitemShut
  {NoStop}%
\bibitem [{\citenamefont {Rivas}\ \emph {et~al.}(2014)\citenamefont {Rivas},
  \citenamefont {Huelga},\ and\ \citenamefont {Plenio}}]{Rivas2014}%
  \BibitemOpen
  \bibfield  {author} {\bibinfo {author} {\bibfnamefont {{\'A}.}~\bibnamefont
  {Rivas}}, \bibinfo {author} {\bibfnamefont {S.~F.}\ \bibnamefont {Huelga}}, \
  and\ \bibinfo {author} {\bibfnamefont {M.~B.}\ \bibnamefont {Plenio}},\
  }\bibfield  {title} {\enquote {\bibinfo {title} {Quantum non-{Markovianity}:
  characterization, quantification and detection},}\ }\href
  {http://stacks.iop.org/0034-4885/77/i=9/a=094001} {\bibfield  {journal}
  {\bibinfo  {journal} {Reports on Progress in Physics}\ }\textbf {\bibinfo
  {volume} {77}},\ \bibinfo {pages} {094001} (\bibinfo {year}
  {2014})}\BibitemShut {NoStop}%
\end{thebibliography}

%

\section{Appendix}%\label{sec:appendix}
In this section we provide an example of {how the code works for a particular process matrix}% a process matrix an input to the code
, and how the different levels of causal information are extracted. In our example we have four parties $\{1,2,3,4\}$, with dimensions 
\[dim = 
  \left[ {\begin{array}{cc}
   d_{1_I} & d_{1_O} \\
   d_{2_I} & d_{2_O} \\
     d_{1_I} & d_{1_O} \\
   d_{2_I} & d_{2_O} \\
  \end{array} } \right]= 
   \left[ {\begin{array}{cc}
   2 & 4 \\
   2 & 8 \\
    2 & 2 \\
   2 & 4 \\
  \end{array} } \right].
\]
Party $1$ has two output subsystems with a dimension of $2$ each, and party $2$ has three output subsystems with a dimension of $2$ each, denoted as $\text{subdim}\{1\} = [2\ 2]$, $\text{subdim}\{2\} = [2\ 2\ 2]$. The process matrix is of the following form
\begin{gather}
W^{1_I1_O2_I2_O3_I3_O4_I4_O}_{input} = \nonumber \\
\rho^{3_I}\otimes T^{3_O1_I}\otimes  T^{1_{O_1}2_I}
\otimes T^{2_{O_3}1_{O_2}4_I}\otimes \id^{2_{O_1}2_{O_2}4_O},
\end{gather}
where $\rho$ is some input state for party $3$, and $1_{O_i}$ and $2_{O_i}$ denote the $i$-th output subsystem of party $1$ and $2$ respectively. Note that the above form of the input process matrix to the code is of course not known in advance. We remind that the input to the code is the above matrix $dim$, the arrays $\text{subdim}\{1\}$, $\text{subdim}\{2\}$ and the process matrix $W_{input}$, in its numerical form, in which the systems are ordered as $1_I1_{O_1}1_{O_2}2_I2_{O_1}2_{O_2}2_{O_3}3_I3_O4_I4_O$.  In the following, we describe the calculations that take place. {The} various procedures can be grouped into three stages.
\\

\emph{Stage 1 - tracing out the open output subsystems.} In this stage, the code looks at the elements subdim\{$X$\}. In our example, $X = 1,2$. Knowing that these parties have output subsystems, it {checks} if those are open---if on the process matrix there is identity on those subsystems. To do that, the code {checks} the following equality for each output subsystem
\be
\tilde{\id} ^{A_{O_i}} \otimes \Tr_{A_{O_I}} W_{input}= W_{input}.
\label{eq:Apcon1}
\ee

The code displays on the command window (see Figure~\ref{fig:command}) the output subsystems for which this constraint is satisfied, and traces {it} out from the process matrix. In our example, it outputs ``There are open subsystems: 1 of party 2 of dimension 2'' and ``2 of party 2 of dimension 2''. The remaining process matrix is now 
%\comment{The different process matrices at different stages should be distinguished, e.g.\ $W_{\textnormal{input}}$, $\bar{W}$, $\tilde{W}$, $W_{(1)}$, or similar.}
\ba
W^{1_I1_O2_I2_O3_I3_O4_I4_O} = \nonumber \\
 \rho^{3_I}\otimes T^{3_O1_I}\otimes  T^{1_{O_1}2_I}
\otimes T^{2_{O_3}1_{O_2}4_I}\otimes \id^{4_O},
\label{eq:Wform}
\ea
and will be used as the input process matrix for the rest of the code. 
\\

\emph{Stage 2 - checking if W is causally ordered.} In this stage, the maximal {non-signaling} sets are established, as well as their causal order. To establish the `last set', which is the set of parties that have no outgoing arrow, the code checks the constraint
\be
\tilde{\id} ^{A_O} \otimes \Tr_{A_O} W= W,
\label{eq:Acon1}
\ee
for all parties $A= \{1,2,3,4\}$. The set of parties that satisfy this constraint comprise of the `last set'. To establish the next set, the last set is traced out from the process matrix and the remaining process matrix undergoes the same above constraint for the remaining parties. {In this way,}
all the maximal sets are established, together with their causal order. If the code completes this task with all the parties grouped into maximal sets, then the process matrix is causally ordered. In our example the maximal sets and their causal order are: $\{3\}\prec\{1\}\prec\{2\}\prec\{4\}$. {This is shown on the command window as (see also Figure~\ref{fig:command})}
\[
\texttt{the\_sets = }
   \left[ {\begin{array}{cc}
  4 \\
   2 \\
    1 \\
  3 \\
  \end{array} } \right].
\]
\\

\emph{Stage 3 - causal discovery and Markovianity.} In this stage the code discovers the causal arrows that connect the parties. Once the all the causal arrows have been found, it checks if the process is Markovian. If it is, it outputs the DAG corresponding to the process. If it is not Markovian, then the discovered causal arrows are not reliable and hence a DAG is not provided. 
Now let us see how the code goes about discovering the causal arrows. The causal arrows are between an input system of a party, say $A$ and an output system or subsystem of another party, say $B$. This is done by the following two constraints for output system or subsystem respectively
\be
\tilde{\id}^{B_O} \otimes \Tr_{B_O} (\Tr_{A_I} W) = \Tr_{A_I} W,
\label{eq:Acon2}
\ee
\be
\tilde{\id}^{B_{O_i}} \otimes \Tr_{B_{O_i}} (\Tr_{A_I} W) = \Tr_{A_I} W.
\label{eq:Acon2i}
\ee
To check whether this constraint is satisfied, the code must check each pair of [input system - output system] or [input system - output subsystem] individually. An input system can be involved with more than one causal arrow, but an output system or subsystem can be involved with only one causal arrow. Hence, once an output system or subsystem has been associated with a causal arrow, it is not checked again in the rest of the code. The code outputs on the command window the causal arrows found. In our example that would be ``Link from subsystem 3 of party 2 to party 4.'', ``Link from subsystem 1 of party 1 to party 2.'', ``Link from party 3 to party 1.'', ``Link from subsystem 2 of party 1 to party 4.'', as is shown on Figure~\ref{fig:command}.

Now the code proceeds with the Markovianity check. This involves constructing a process matrix that is Markovian with respect to the found DAG{; specifically, a} matrix composed out of input states for the first parties, channels for the found causal arrows and identity matrices for the last parties. For the first parties, it extracts from the process matrix their input states. In our example, party 3 is first and its input state is extracted from the process matrix by tracing out all the other systems
\be
\rho^{3_I} = \Tr_{\widetilde{3_I}}W,
\ee
where $\widetilde{3_I}$ denotes the space of input and output systems that is complementary to $3_I$. To extract the channels, the code similarly traces out all systems except the ones involved in the channels. Note that for an input system that is involved with many arrows the corresponding channel would be represented from the parent space of the party---all the systems that have an arrow to that party---to the input of the party. In our example, there are the simple channels from systems $3_O$ to $1_I$, from $1_{O_1}$ to $2_I$, and from \{$1_{O_2}$,$2_{O_3}$\} to $4_I$. The corresponding channels are
\ba
T^{3_O1_I} = \Tr_{\widetilde{3_O1_I}} W\\
T^{1_{O_1}2_I} = \Tr_{\widetilde{1_{O_1}2_I}} W\\
T^{I_{O_2}2_{O_3}4_I} = \Tr_{\widetilde{I_{O_2}2_{O_3}4_I}} W
\ea
The code is then adding identities to the output systems of the last parties. In our example we have $\id^{4_O}$. Finally, the code {constructs}
the following test matrix 
\ba
W_{\textnormal{test}}^{1_I1_O2_I2_O3_I3_O4_I4_O} = \nonumber \\
\rho^{3_I}\otimes T^{3_O1_I}\otimes  T^{1_{O_1}2_I}
\otimes T^{2_{O_3}1_{O_2}4_I}\otimes \id^{4_O}.
\label{eq:Wform2}
\ea
After rearranging the systems in the order of the original process matrix, that is $[1_I1_O2_I2_O3_I3_O4_I4_O]$, the code tests if $W_\textnormal{test }= W$. If this is true, which it is in our example, the code outputs on the command window ``the process is Markovian'' and outputs the DAG corresponding to the found causal arrows, shown in Figure~\ref{fig:dag}. 
\\
\bef
\includegraphics[width=\linewidth]{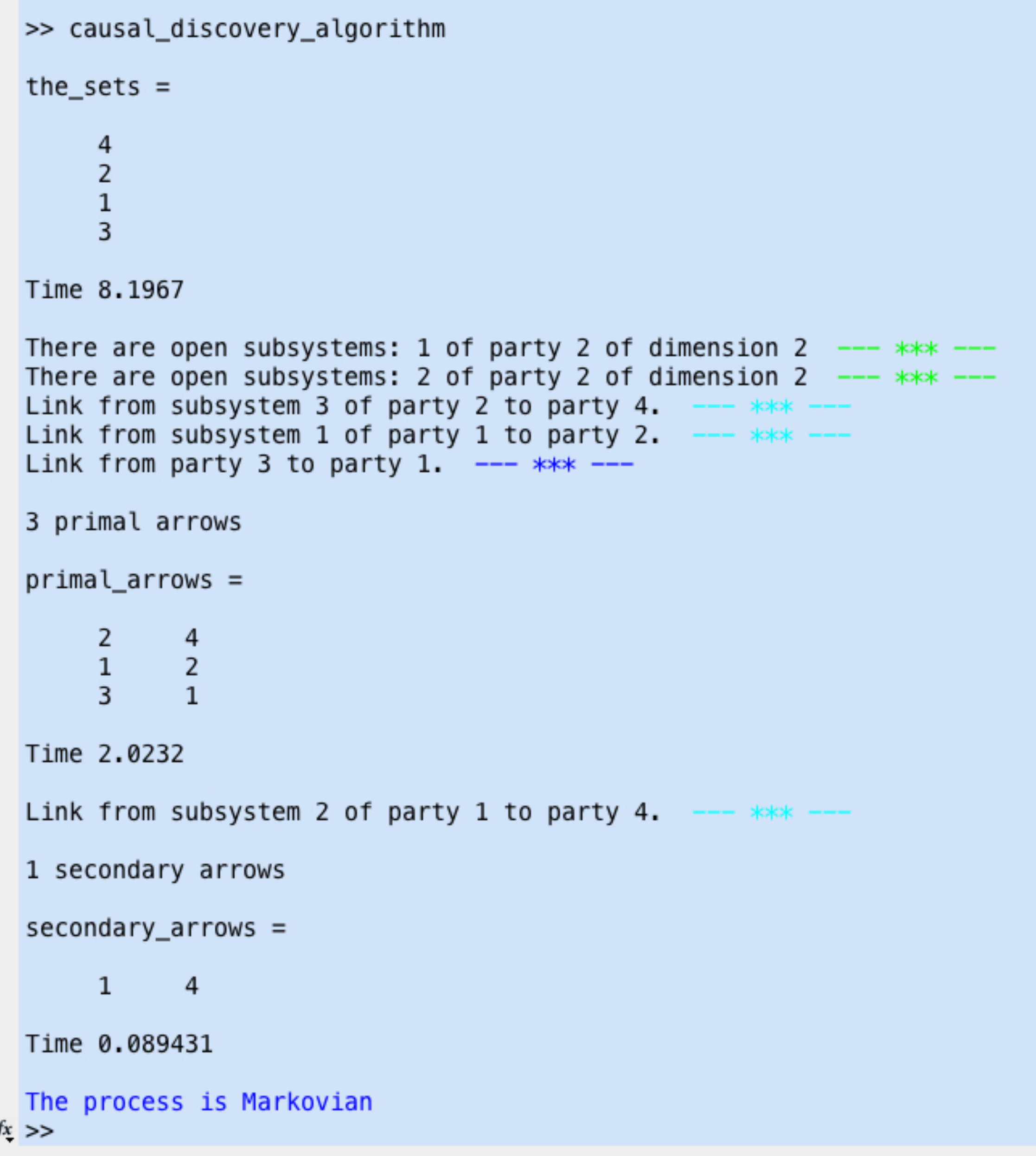}
\caption[]{\textbf{Output of command window}: The command window, for the given example, showing the output of the code regarding the maximal non-signaling sets, the open subsystems and the causal arrows. Primal\_arrows refers the the causal arrows from successive maximal non-signaling sets and secondary\_arrows refers to all the {other} %rest 
causal arrows. Time refers to the time that lapsed to evaluate the step just above.}
\label{fig:command}
\eef
\\
\bef
\includegraphics[width=.65\linewidth]{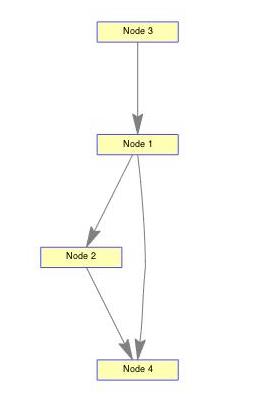}
\caption[]{\textbf{Output DAG}: The DAG that the code outputs for the given example.}
\label{fig:dag}
\eef

In the given repository~\cite{discovery_code2017}, we provide the code presented in this paper, written in MatLab, together with the set of necessary functions. We also provide a code written in Mathematica, where valid process matrices of arbitrary causal structures can be generated, given the number of parties. These process matrices can be used as examples of input to the code. Finally, we provide a Manual on how to use both codes.

\end{document}